# Halogen Chains with One-Dimensional Semi-Metallic Electronic Structure and Peierls Physics in Polymorphs of $Na_4X_5$ (X = I, Br, Cl) Compounds

Yuqing Yin[1,2], Nityasagar Jena[2], Florian Knoop[2], Gaston Garbarino[3], Ferenc Tasnádi[2], Florian Trybel[2], Sergei I. Simak[2,4], Mikhail I. Katsnelson[5,6], Alena Aslandukova[7], Andrey Aslandukov[7], Xiang Li[3], Anaëlle Antunes[8], Fariia Iasmin Akbar[7], Konstantin Glazyrin[9], Haixing Fang[3], James A. D. Ball[3], Natalia Dubrovinskaia[1], Leonid Dubrovinsky[8]*, Igor A. Abrikosov[2]*

**Affiliations:**

[1]Material Physics and Technology at Extreme Conditions, Laboratory of Crystallography, University of Bayreuth; Bayreuth, 95440, Germany.
[2]Department of Physics, Chemistry and Biology (IFM), Linköping University; Linköping, SE-581 83, Sweden.
[3]ESRF, The European Synchrotron; 71 Avenue des Martyrs, CS40220, 38043 Grenoble Cedex 9, France.
[4]Department of Physics and Astronomy, Uppsala University; Box 516, SE-751 20 Uppsala, Sweden.
[5]WISE-Wallenberg Initiative in Materials Science, Uppsala University; Box 516, SE-751 20 Uppsala, Sweden.
[6]Institute for Molecules and Materials, Radboud University; Nijmegen, 6500 HC, Netherlands.
[7]Institute of Inorganic and Analytical Chemistry, Goethe University Frankfurt; Max-von-Laue-Straße 7, Frankfurt, 60438, Germany.
[8]Bayerisches Geoinstitut, University of Bayreuth; Bayreuth, 95440, Germany.
[9]Deutsches Elektronen-Synchrotron DESY; Notkestr. 85, Hamburg, 22607, Germany.

*Corresponding author. Email: Leonid.Dubrovinsky@uni-bayreuth.de; igor.abrikosov@liu.se

**Abstract:** Since the pioneering works of Peierls, one-dimensional materials have attracted great attention. Still, the synthesis of truly monoatomic chains remains elusive. In this study, we explore a novel path of experimental synthesis of monoatomic one-dimensional chains by their chemical stabilization in ionic compounds. We demonstrate that in synthesized at high pressure sodium halides $Na_4X_5$ (X = I, Br, Cl) with *hP*18 $Ga_4Ti_5$-type structures, transfer of valence electrons from cations to anions leads to the formation of halogen chains connected with other atoms only by ionic interaction and having one-dimensional electronic structure. The Peierls physics in the systems is confirmed by theoretical calculations, newly synthesized incommensurately modulated *i*-*hP*18-$Na_4X_5$ (X = I, Br, Cl) compounds, as well as by the discovered *hP*36 phases of $Na_4Cl_5$ and $Na_4Br_5$.

The concept of a one-dimensional (1D) chain of atoms is familiar to every physicist, chemist, or materials scientist. It serves as a textbook model to explain the idea of translational lattice periodicity, lattice vibrations or electronic structure relations of a simple periodic system.(*1*) 1D materials are of great interest due to the possibility to study spin-charge separation, charge and spin-density waves, and topological spin excitations.(*2*) Most solid-state systems we deal with in practice are three-dimensional (3D). The so-called low dimensional systems have attracted a great deal of attention in recent years, as they can exhibit numerous exotic electronic, magnetic, and transport phenomena,(*3, 4*) and models for 2D—and especially 1D—systems are often easier to formulate and to treat, allowing exact or numerical solutions of the governing equations. Notably, in the systems with 1D electronic structures, the conventional Fermi liquid theory breaks down, and electrons are better described by the Tomonaga-Luttinger liquid model.(*5, 6*) In the presence of a constriction or quantum point contact, conductance in 1D systems is quantized (*7*) and there are topologically protected states which are robust against perturbations.(*8*)

While systems with 1D electronic structures are highly attractive for fundamental studies and may have high potential for advanced technological applications, their existence in the real world is considered highly unfeasible. Unlike 2D materials, such as graphene, that can be macroscopically stabilized, as a free-standing system or over a suitable substrate (e.g., hexagonal boron nitride, h-BN) (*9*), realizing isolated 1D structures is much more difficult in practice. Indeed, both 1D and 2D crystals are potentially unstable due to strong fluctuations associated with soft phonon modes. However, in two dimensions the corresponding contribution to mean-square interatomic displacement diverges only logarithmically. This means that short-range crystalline order survives till large distances, and long-range order can be stabilized by already relatively weak interaction with a substrate.(*10*) In the 1D case, we deal with a much stronger, power-law divergence, which makes stabilization problematic. Importantly, strictly 1D systems are intrinsically susceptible to structural instabilities, such as the Peierls transition, that opens up an energy gap in their electronic structure at the Fermi level, thus suppressing metallic conductivity.(*11*)

In the absence of systems with truly 1D electronic properties, several quasi-1D (q1D) materials have been explored as proxies.(*12*) Among them, for example, are nanowires of sub-nanometer diameters stabilized on substrates or nanotubes (including carbon nanotubes) that demonstrate signs of physical properties expected for 1D electron systems.(*13, 14*) Another important class of q1D systems are molecular crystals with crystal structures that have highly anisotropic motifs along one crystallographic direction, such as tetrathiofulvalinium tetracyanoquinodimethane (TTF-TCNQ).(*15, 16*) Due to the localization of the "movement" of electrons along (almost) 1D, q1D materials have been reported to have variety of emergent physical properties, such as $\pi$-electron based covalent antiferromagnetism in $Mn_2Hg_5$,(*17*) nearly ferromagnetic spin-triplet superconductivity in $UTe_2$ and $K_2Cr_3As_3$,(*18, 19*) and q1D superconductivity in $Na_{2-\delta}Mo_6Se_6$.(*20*)

In this study, we suggest that an ideal 1D electron system can be realized in a crystal comprising atomic chains hosting fully delocalized electrons, where the 1D electronic subsystem is effectively decoupled from the surrounding structure. In such a scenario, the 1D atomic chain (or chains) is structurally embedded within a 3D crystalline host yet maintaining isolated 1D electronic behavior. Based on our suggestion, we explore a novel path of experimental synthesis of monoatomic 1D chains by their chemical stabilization in predominantly ionic compounds. We demonstrate that in sodium halides with chemical composition $Na_4X_5$ (X = I, Br, Cl) synthesized at high pressure (HP), transfer of valence electrons from cations to anions leads to the formation of linear halogen chains with delocalized one-dimensional electronic structure, decoupled from the electron subsystem of other atoms. Experimental structural studies and theoretical analysis show clear signatures of Peierls physics in these materials. We demonstrate that at zero temperature, in agreement with what is expected for Peierls materials,(*11*) the electron-phonon interactions destabilize the linear monoatomic metallic chains of halogen anions, leading to opening band gaps in the electronic structure. However, at finite temperature, anharmonic effects of lattice vibrations lead to stabilization of regular monoatomic chains of equidistant halogen anions, as experimentally observed in the *hP*18-$Na_4X_5$ polymorphs, closing the band gaps, and making the one-dimensional chains semi-metallic.

## Synthesis of *hP*18 $Na_4I_5$ with 1D halogen chains preserved down to ambient pressure

The synthesis of *hP*18 $Na_4X_5$ (X = Br, Cl) compounds (space group #193, $P6_3/mcm$) at 48-73 GPa has been reported by some of the authors of this paper in Ref. (*21*). In fact, these compounds adopt the $Ga_4Ti_5$-type structure.(*22*) Assuming that an increase of the atomic number of the halogen leads to a decrease in the synthesis pressure,(*23*) we studied the Na–I system and successfully synthesized a previously unreported iodine compound, *hP*18-$Na_4I_5$, at pressures of about 20 GPa. The compound was quenched to ambient conditions in an inert atmosphere, and we demonstrated that its synthesis is possible and we successfully achieved the synthesis at pressure as low as 3(2) GPa. A series of experiments, which led to the synthesis are described below.

Upon laser heating (LH) of a mixture of NaI and $CI_4$ at ~2300 K at 20(2) GPa (DAC #1) and 24(2) GPa (DAC #2) (see Supplementary Methods and Table S1), we observed formation of a yet unknown phase with the composition $Na_4I_5$ as revealed by *in situ* single-crystal X-ray diffraction (SCXRD, see Supplementary Methods). This phase, *hP*18-$Na_4I_5$ (Fig. 1b, c), was identified as isostructural to *hP*18-$Na_4Cl_5$ and *hP*18-$Na_4Br_5$ reported earlier.(*21*) In the structure of *hP*18-$Na_4I_5$ (Fig. S1) sodium atoms form two different polyhedra – distorted $Na1I2_6$ octahedra and $Na2I_9$ distorted capped square antiprisms. Na2 is coordinated by both the I1 and I2 atoms. Na1:Na2 ratio is 1:3. The face-sharing $Na1I2_6$ octahedra form columns along the *c* direction, and $Na2I_9$ polyhedra fill the space between the columns (Fig. S1). The Na-I distances in sodium capped square antiprisms (Na2-I) are significantly longer than in distorted octahedra (Na1-I2) (for example, average distances of Na2-I and Na1-I at 10(2) GPa and ~130 K are 3.13 and 2.85 Å, respectively).

The behavior of *hP*18-$Na_4I_5$ was investigated at variable pressures and temperatures (down to ~20 K, Table S1-S2). High-quality SCXRD data of *hP*18-$Na_4I_5$ were collected on decompression at room temperature from 24(2) GPa to 10(2) GPa. At lower pressures, the quality of single crystals degraded making further SCXRD measurements infeasible. However, powder XRD (PXRD) was measured at the lowest pressure point of 6(2) GPa (DAC #1). Considering the known sensitivity of polyhalogenides to moisture, the samples were fully decompressed ("recovered") to ambient pressure by opening the DAC in an inert atmosphere (in a glove bag or a glove box). After subsequently re-closing the DAC in the same bag/box following recovery, the pressures were found to be 2(2) GPa and 3(2) GPa in DAC #1 and DAC #2, respectively. A comparison of PXRD patterns taken upon decompression (6(2) GPa) and after recovery (3(2) GPa) (Fig. S2a) provides evidence that *hP*18-$Na_4I_5$ compound was preserved down to ambient pressure in an inert atmosphere.

## Incommensurately modulated *i-hP*18-$Na_4I_5$

DAC #2, containing the $hP18$-$Na_4I_5$ sample held at 3(2) GPa, was stored for more than 35 days and subsequently subjected to additional laser annealing at ~2300 K. After the annealing, the material recrystallized and SCXRD data revealed two phases. One of them was the original $hP18$-$Na_4I_5$ phase. At the same time, the detailed analysis of reconstructed reciprocal space images and satellite peak intensity profiles for several crystalline domains revealed superlattice reflections, and the second phase was identified as incommensurately modulated *i*-*hP*18-$Na_4I_5$ (see Supplementary Discussion 1). The superlattice reflections could be indexed with a modulation wavevector $\mathbf{q} = 0.4858\mathbf{c}^*$, with $\mathbf{c}^*$ being a reciprocal lattice vector of the unit cell. Therefore, the structural model was refined as incommensurately modulated *i*-*hP*18-$Na_4I_5$, described by the (3+1)-dimensional superspace group $P6_3/mcm(00\gamma)00ss$ (Fig. 1e). The refined structural parameters at 3(2) GPa are $a$ = 10.111(2) Å, $c$ = 6.5847(5) Å, $\gamma$ = 0.4858(8) (Table S3). Displacive modulations of the atomic positions ($u_i(\overline{x_4})$ for $i = x, y, z$) are described by a truncated Fourier series, considering only the 1st order harmonics, given by: $u_i(\overline{x_4}) = A_1(i)\cos(2\pi\overline{x_4})$, where $\overline{x_4} = t + \boldsymbol{q}\overline{\boldsymbol{x}}$. For I1 atoms, the modulation amplitude is $A_1(z)$ = 0.0443(3). Minor modulation amplitudes are observed for I2 and Na2 atoms, with $A_1(x)$ = 0.00282(5) and $A_2(y)$ = 0.00563(10) for I2, and $A_1(x)$ = 0.017(2) and $A_2(y)$ = 0.0087(10) for Na2. In total, 957 independent reflections were collected with 392 main and 565 satellites of the first order. The final agreement factors converged to the values of around 7.0% ($R_1$) for all observed reflections ($I > 3\sigma(I)$) and around 9.8% ($wR_2$) for all reflections (Table S3).

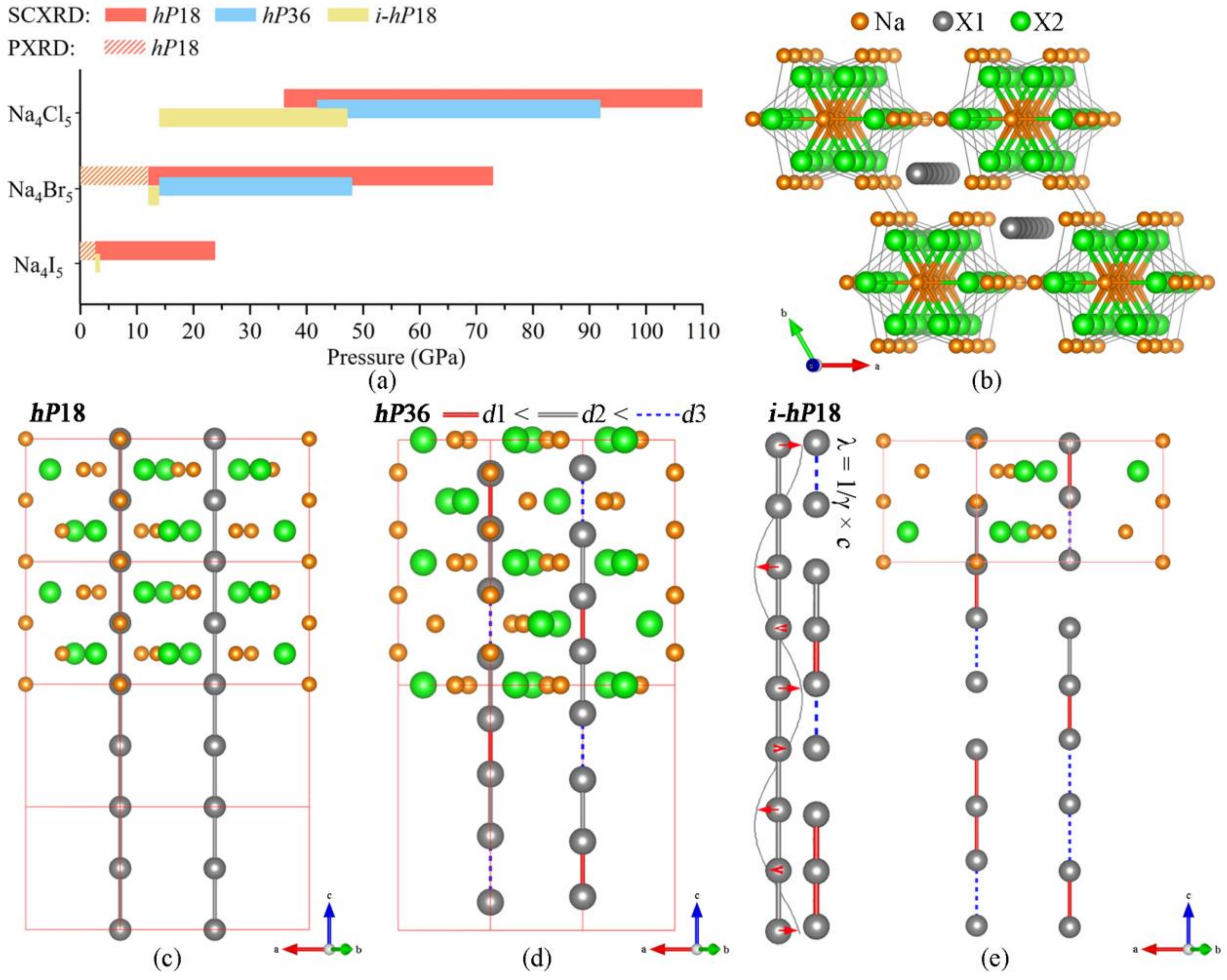

**Figure 1. Occurrence and crystal structures of $Na_4X_5$ (X = I, Br, Cl) compounds.** (a) Schematic diagram presenting pressure intervals in which various $Na_4X_5$ polymorphs were experimentally observed at room temperature (298 K). Color code relates to the structure of polymorphs notated by Pearson symbols. Full color corresponds to the structural data obtained using SCXRD; hatching marks the data obtained by PXRD. Different compounds are noted to the left. The letter *i*, added to the Pearson symbol (*i-hP*18), refers to incommensurately modulated structures with the space group $P6_3/mcm(00\gamma)00ss$. (b) Ball-and-stick model for the *hP*18 polymorph in perspective view. (c), (d), and (e) are ball models of the *hP*18, *hP*36, and *i-hP*18 polymorphs, correspondingly, as viewed along the [120] direction; the common structural motif – linear X1 chains characteristic for all three polymorphs – is highlighted by grey straight lines. Lattices of the periodic basic structures are indicated by red thin lines. Na atoms are orange, X1 atoms are grey, and X2 atoms are green. Na and X2 atoms are omitted at the bottom to highlight the X1 chains. Incommensurate longitudinal modulation in (e) is visualized by thin sinusoidal line along the X1 chain. The variation of the $\gamma$ component of the modulation wavevector – which fixes the wavelength of the modulation –is plotted in Fig. S3 as a function of pressure.

## New *hP*36-$Na_4Cl_5$ and *hP*36-$Na_4Br_5$ polymorphs: further signatures of Peierls physics

Unexpected finding of the *i-hP*18-$Na_4I_5$ phase motivated us to investigate the Na-Cl and Na-Br systems in a wider pressure range (up to ~110 GPa and ~73 GPa respectively, Fig. 1a, Table S1) than in our earlier study.(*21*) Special attention was paid to the appearance of possible superlattice reflections. As a result, we found incommensurate *i-hP*18-$Na_4Cl_5$ and *i-hP*18-$Na_4Br_5$ phases: the former occurring in a broad pressure range of ~14-47 GPa and the latter – in a narrow interval of ~12-14 GPa (Fig. 1a). Additionally, apart from the already reported *hP*18-$Na_4Cl_5$ and *hP*18-$Na_4Br_5$ phases,(*21*) new *hP*36-$Na_4Cl_5$ and *hP*36-$Na_4Br_5$ polymorphs (space group #190, $P\bar{6}2c$, Fig. 1d) were found in broad pressure ranges (see Fig. 1a and Table S1 for details of the synthesis). The *hP*18 and *hP*36 phases co-exist in wide but limited pressure ranges (Fig. 1a).

The structures of *i-hP*18, *hP*18, and *hP*36 polymorphs of $Na_4X_5$ (X = I, Br, Cl) compounds are closely related (see Fig. 1b-e, Tables S2-S7). The ball-and-stick model of the *hP*18 structure (Fig. 1b) highlights rows of Na and X atoms along the *c* direction. The ball models of all three structures projected along the [120] direction (Fig. 1c-e) depict their differences and similarities. The unit cell parameter *c* of *hP*36 phases is about two times larger than that of *hP*18. In *hP*18 structures the distances between Na1 atoms in the rows running along the $6_3$ [00z] axis are very short, especially in $Na_4Cl_5$, where they are shorter than Na-Na contacts in metallic *bcc*-Na at corresponding pressures (*21*). X1 halogen atoms align along the $\bar{3}$ axis forming 1D chains. These chains (Fig 1b) appear to be the most distinct structural elements, which define the differences between all three types of $Na_4X_5$ structures. Whereas in *hP*18 polymorphs all X1 atoms are equidistant (Fig 1c), in *hP*36 and *i-hP*18 polymorphs (Fig. 1d-e) the X1-X1 distances are modulated.

The *i-hP*18-$Na_4X_5$ phases emerge at lower pressures as a result of incommensurate modulation of X1-X1 distances along the 1D chains (Fig. 1e). Remarkably, the variation of the distances are very significant (Fig. S4): for example, I–I contacts are varying from 2.913(3) to 3.672(3) Å in *i-hP*18-$Na_4I_5$ at 3(2) GPa, or from 2.184(6) to 2.972(6) Å in *i-hP*18-$Na_4Cl_5$ at 14(2) GPa (for comparison, typical variations of interatomic distances in q1D materials with charge density waves due to a Peierls instability are about 0.025-0.2 Å (*24*)).

In $hP36$-$Na_4X_5$ halogen atoms form linear tetrads [$Cl_4$] with shorter interatomic contact between inner atoms (d1) and a longer one between inner and outer atoms (d2). The tetrads are separated at a distance d3, so that d1 < d2 < d3 (Fig. 1d). For example, in $hP36$-$Na_4Cl_5$ at 60(4) GPa, the Cl-Cl distances within the [$Cl_4$] units are d2 = 2.353(3) – d1 = 2.126(4) – d2 = 2.353(3) Å, and d3 = 2.566(4) Å between them. Considering that the shortest (d1) distances in [$X_4$] tetrads in $hP36$-$Na_4Cl_5$ are comparable with intramolecular distances in molecular halogen crystals at corresponding pressures, and d2 distances are similar to those in halogen chains of $[X_2]^-_\infty$ in $hP18$-$Na_4X_5$,(*21*) the formal charges in $hP36$ phases can be assigned as $Na^+_8X^-_6[X_4]^{2-}$. The "-2" formal charge of [$X_4$] groups is in good agreement with that proposed for polyiodide $[I_4]^{2-}$ ions.(*25*)

The synthesis conditions for $Na_4X_5$ (X = I, Br, Cl) are summarized in Fig. 1a, which visualizes experimentally derived phase relations. As seen, the $hP18$ $Na_4X_5$ compounds, featuring ideal linear chains of the equidistant halogen atoms, may coexist with the $hP36$ phases, in which the chains are distorted due to the formation of $[X_4]^{2-}$ triads, and *i*-$hP18$-$Na_4I_5$ structures with the incommensurate longitudinal modulation along the chains. The remarkable co-existence of $hP18$, *i*-$hP18$, and $hP36$ phases in the same samples suggests that there are first order phase transitions between them. According to our observations, $hP18$-$Na_4I_5$ phase remains until 20 K upon its cooling at 21(2) GPa (Fig. S5a). The structural information derived from the experiment indicates that the halogen chains in the synthesized compounds bear clear signatures of the Peierls distortions typical for true 1D systems.

**Phase stability of polymorphs of $Na_4X_5$ (X = I, Br, Cl) compounds**

To confirm the Peierls physics in our experimentally studied systems, we performed their theoretical analysis. We also tried to identify the origin of the effects characteristic for a 1D material in the discovered 3D compounds. Starting with the verification task, we carried out the Density Functional Theory (DFT) calculations at $T = 0$ K (see Supplementary Methods) of structural properties of all $hP18$ and $hP36$ $Na_4X_5$ compounds in pressure intervals corresponding to the synthesis pressure (Table S8). We found good agreement of the theoretically calculated fully relaxed structures and those experimentally observed (Table S8). Theoretical equations of states were calculated for $hP18$ and $hP36$ polymorphs of all $Na_4X_5$ (X = I, Br, Cl) phases, although $hP36$-$Na_4I_5$ was not observed experimentally. They appeared to be in reasonable agreement with the experiment for $Na_4I_5$, $Na_4Cl_5$, and $Na_4Br_5$ (Fig. S5b-d, and Table S9). The enthalpy differences between $hP18$ and $hP36$ $Na_4X_5$ phases, calculated at zero temperature (Fig. S6), are in accord with the experimental observation that the former are favored by compression and by increase of the atomic masses of halogen atoms (Fig. 1a). Phonon dispersion relations calculated for the $hP18$ compounds at $T = 300$ K (Fig. S7) demonstrate that they are dynamically stable in their experimentally determined pressure intervals, with the exception of the lowest pressure: $hP18$-$Na_4Cl_5$ and $hP18$-$Na_4I_5$ according to the calculations were dynamically unstable at about 50 GPa and 3 GPa, respectively, although they were observed in experiments at those pressures and below (Fig. S8). The pressures at which phonon soft modes appear in our calculations for $hP18$-$Na_4X_5$ compounds near the A-point in the Brillouin zone correspond to the experimentally observed incommensurately modulated structures with the space group $P6_3/mcm(00\gamma)00ss$ at lower pressures (Fig. S7). Furthermore, the phonon dispersion relations calculated for $hP36$-$Na_4X_5$ (X = Cl, Br) show that they are dynamically stable already in the harmonic approximation, that is at $T = 0$ K (Fig. S9). Thus, we conclude that our DFT calculations (see Supplementary Methods) well reproduce structural and compressional properties of the materials of interest.

This allows us to proceed further with the analysis of the chemical nature of the new phases, which ought to be the origin of the observed experimental signatures of the Peierls physics in the studied crystals.

**Electronic isolation of the halogen chains due to charge transfer effects**

A very important point for the analysis is the fact that calculated X-X distances in the 1D halogen chains are in reasonable agreement with experiment (Fig. S4). Thus, DFT calculations carried out for the *hP*36 $Na_4I_5$, $Na_4Br_5$ and $Na_4Cl_5$ at $T = 0$ K confirm the formation of the $X_4^{2-}$ units in these phases (Figs. S4, S10b). For *hP*36-$Na_4X_5$ there are three independent positions of Na atoms per unit cell, and all of them have about the same Bader charges approximately equal to the Bader charge of Na in B1 or B2 structured NaX at the same pressure (Table S10). Unit cell of *hP*36-$Na_4X_5$ contains four crystallographically distinct X atoms: 12 X individual atoms have Bader charges opposite to those of Na, and 8 atoms form two symmetrically identical [$X_4$] groups. In each [$X_4$] group two inner (spaced at a shorter distance) halogens have absolute values of the Bader charges ~40% of that of Na, two outer halogens – ~60% of that of Na, making the absolute value of the total Bader charge of the [$X_4$] unit of about twice of sodium. Thus, DFT calculations support the derived above crystal-chemical formula of *hP*36-$Na_4X_5$ compounds as $Na^+_8X^-_6[X_4]^{2-}$, showing that within $[X_4]^{2-}$ anions outer atoms have higher charges than inner atoms (similar to that proposed for $[I_4]^{2-}$ ions (*25*)), and suggesting ionic interaction between $Na^+$, $X^-$, and $[X_4]^{2-}$ constituents of the phases.

Calculated Bader charges of the *hP*18-$Na_4X_5$ compounds (Table S10) support the picture of formal charge distributions in the compounds based on qualitative crystal chemical analysis presented above. Indeed, for all the *hP*18-$Na_4X_5$ phases Bader charges of both structurally independent Na at Wyckoff sites 2*b* (Na1) and 6*g* (Na2) are like those in the corresponding NaX salts, suggesting that sodium atoms transfer their electrons to halogen atoms forming a positively charged closed-shell ions. Contrary, halogens are clearly divided into two types (Table S10). First, X2 (in 6*g* Wyckoff sites) atoms have Bader charges about the same by absolute value, but opposite to Na. Like X atoms in NaX salts, X2 atoms in *hP*18-$Na_4X_5$ form negatively charged closed shell ions. Second, X1 (in 4*d* Wyckoff sites) have the Bader charges that are about ½ of X2 ions charge. If we accept a formal charge of closed shell sodium ion as +1, the formula of the compounds will look like *hP*18-$Na^+_4X^-_3[X_2]^-_\infty$ (where $[X_2]^-_\infty$ means $^1_\infty[X_2]^-$ 1D polyanion) in accordance with crystal-chemical consideration. The presented picture of the distributions of charges is additionally supported by the calculated charge density isosurfaces (Fig. S10) that suggest ionic bonding between $Na^+$, $X2^-$, and polyanion $[X1_2]^-_\infty$.

The most remarkable features of the studied $Na_4X_5$ compounds which emerge from the crystal-chemical analysis can now be summarized as follows. The 1D chains formed by X atoms (which do not have enough electrons to form closed shells ions), either polyanions $[X1_2]^-_\infty$ in *hP*18 phases or $X_4^{2-}$ tetrads in *hP*36 phases, are electronically isolated from the environment, as they are embedded into purely ionic cages formed by the $Na^+$ ions and remaining $X^-$ ions (Figs. 1b, S10). The presented picture is additionally confirmed by atom/orbital projected electronic density of states (eDOS) illustrated for *hP*18-$Na_4I_5$ in Fig. S11. We therefore expect that the chains host the 1D electronic structure.

**Electronic structure of the ideal *hP*18-$Na_4X_5$ compounds at *T* = 0 K**

Indeed, our DFT calculations for the ideal (static) phase of *hP*18-$Na_4I_5$ at *T* = 0 K show (Fig. 2a) two (undistinguishable within the figure resolution) cosine-shaped electronic bands along the Cartesian *z*-direction (Γ–A, L–M, and K–H) exhibiting vanishing dispersion in the *x*-*y* plane, as emphasized by the planar Fermi surfaces (Fig. 2b). These features firmly demonstrate the one-dimensional metallic character of $Na_4I_5$ in the ideal *hP*18 crystal structure at *T* = 0 K. The orbital-resolved electronic band structure ("fat bands") shows that the cosine-shaped electronic bands are due to $5p_z$ orbitals of iodine arranged in $[I1_2]^-_\infty$ chains. We note that the flat bands (as shown in red color in Fig. 2a) along Γ-M-K-Γ in fact cross each other at the K point (Fig. S12), however the negligible gradient of the electronic bands at the crossing point underlines the vanishing interaction (electron hopping) between the $[I1_2]^-_\infty$ chains. A very important additional observation is that the bands due to $Na^+$ ions and remaining $I^-$ ions are well separated from each other and are located far away from the Fermi energy, in agreement with the picture of the electronic isolation of the $[I1_2]^-_\infty$ chains from the rest of the environment.

We have constructed the bonding and antibonding Wannier functions of the I1 $5p_z$ orbitals (Fig. 2c) and have reproduced the cosine-shaped bands (Fig. S13), in good agreement with our DFT results in Fig. 2a. The spatial distribution of the Wannier functions, together with the calculated total electronic charge density in Fig. S10, reveal that the iodine I1 atoms along the 1D $[I1_2]^-_\infty$ chains would form conducting atomic wires shielded by the network of ionic $Na^+$ and $I2^-$ ions in the unit cell of *hP*18-$Na_4I_5$. The same qualitative picture of the electronic structure is observed for *hP*18-$Na_4Br_5$ and *hP*18-$Na_4Cl_5$ compounds (Fig. S14). Although the *hP*18 $Ga_4Ti_5$-type structure is adopted by other compounds, our calculations do not show any 1D features in their electronic structures (see Supplementary Discussion 2). This is in line with the fact that, to the best of our knowledge, neither of these compounds has either *hP*36 or *i*-*hP*18 polymorphs. The *i*-*hP*18 phase of $Ba_3ScTe_5$ (*26*) we found in the literature has not been observed in the *hP*18 $Ga_4Ti_5$ structure type (see Fig. SD2. 5 in Supplementary Discussion 2). The strong charge transfer effects leading to the electronic isolation of the halogen chains make the electronic behavior of *hP*18 $Na_4X_5$ compounds unique.

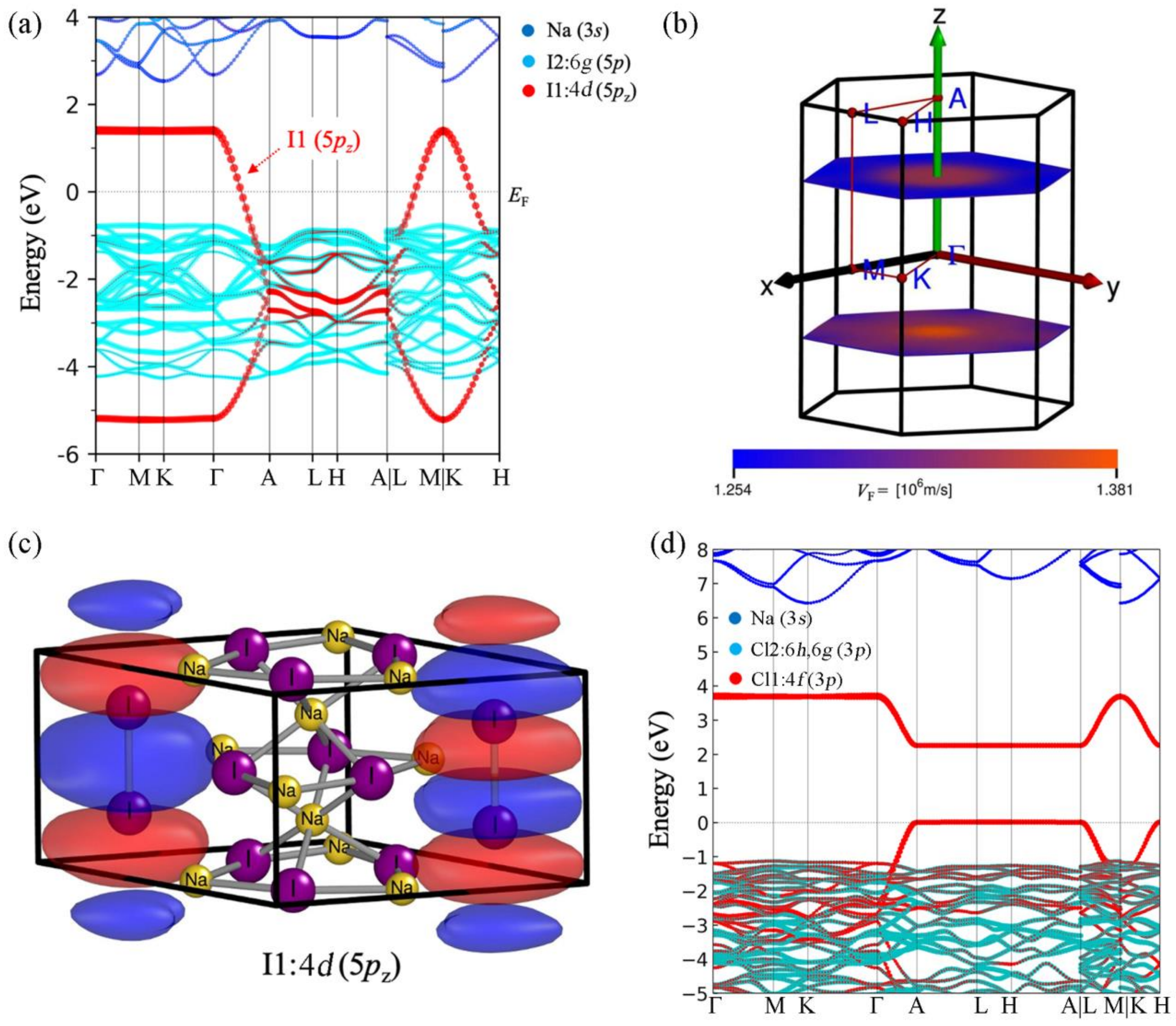

**Figure 2. Electronic structure of $Na_4X_5$ compounds considering ideal (static) crystal structure at $T$ = 0 K.** (a) Orbital projected band structure of *hP*18-$Na_4I_5$ at 5 GPa, the orbital projections to different cation and anion Wyckoff sites have been colored differently for clarity. The I1 atoms in two linear chains in the unit cell at the Wyckoff site 4*d* and the 1D electron density mainly consist of overlapping I1-5$p_z$ orbitals along the linear chain direction. The Fermi level is set to zero and marked with a dashed line. (b) Fermi surface of *hP*18-$Na_4I_5$ at 5 GPa with Fermi velocity ($V_F$) scaled in color. (c) The derived Wannier functions of the I-4*d* (5$p_z$) orbitals that form the 1D electronic bands (as shown in Fig. S13). (d) Band structure of *hP*36-$Na_4Cl_5$ at 50 GPa with band gap $E_g$ = 2.35 eV.

## Peierls physics in polymorphs of $Na_4X_5$ (X = I, Br, Cl) compounds

At this point, we underline that the calculated electronic structure (Fig. 2, S11-S14) has been obtained for the ideal static structure of *hP*18-$Na_4X_5$ at $T$ = 0 K, that is with atoms fixed at their positions and neglecting any effects of atomic vibrations. On the other hand, it is known since the works of Peierls,(*11*) that the metallic state of a 1D system at zero temperature is unstable with respect to the formation of dimers along the chain. Indeed, our calculations of the phonon dispersion relations for *hP*18 $Na_4X_5$ compounds at $T$ = 0 K using the harmonic approximation (see Supplementary Methods) result in a pronounced dynamical instability for the *hP*18 $Na_4Cl_5$, $Na_4Br_5$ and $Na_4I_5$ at the wave vector corresponding to vibrations of the halogen atoms along the chain (Fig. S15). The calculated wave vector for the imaginary frequencies indicating dynamical instability is also in agreement with the formation of earlier unreported *hP*36 phases of $Na_4Br_5$ and $Na_4Cl_5$, as well as *i-hP*18-$Na_4X_5$ phases synthesized in our experiments (Fig. 1 and Figs. S4-S5).

According to the Peierls picture,(*11*) an arbitrarily weak electron-phonon coupling leading to the structural distortions of the 1D chains opens an energy gap in the electronic structure at the Fermi level, reducing the system's total energy by introducing elastic energy through lattice modulation, while simultaneously lowering the kinetic (single-particle electronic) energy. The electronic structure calculations for the *hP*36 $Na_4I_5$, $Na_4Br_5$ and $Na_4Cl_5$ confirm that the band gaps open in all the compounds (see Fig. 2d and Fig. S16), clearly pointing towards the presence of Peierls physics in these materials.

While the ideal 1D chains are unstable at $T = 0$ K, in agreement with our calculations (Fig. S15), the finite temperature effects should lead to their stabilization in the studied systems. At finite temperature, the system should optimize, not the total energy but, rather, free energy, which includes also entropy contribution. The gapless phase has much higher electron entropy than the gapped one (*1*) which leads to metal-insulator transition with the gap closure at some critical temperature.(*12*) This explains the stabilization of the *hP*18 phases as the temperature increases. Note that the width of the band gaps of *hP*36-$Na_4X_5$ also reduces with the pressure, as shown in Fig. S16, confirming its stabilizing effect on the *hP*18 phases with respect to *hP*36 phases seen in our experiments.

We investigated the intriguing interplay between phonons and electrons in *hP*18-$Na_4X_5$ using *ab initio* molecular dynamics (AIMD) simulations accelerated by on-the-fly machine-learned force field method (see Supplementary Methods). As pointed out earlier, the phonon dispersion relations calculated for *hP*18-$Na_4X_5$ (X = I, Br, Cl) compounds at $T = 300$ K confirm their dynamical stability in good agreement with the experimentally determined pressure intervals (Fig. S7, Fig. S8). Fig. S17 shows that phonon dispersion relations and the projected phonon density of states (PhDOS) of the studied systems reveal a notable trend across the halogen series (X = Cl, Br, I). For the Cl-containing compound, the PhDOS of Na and Cl exhibit substantial overlaps, suggesting strong vibrational coupling between the Na and Cl ions, though the soft mode near the A point in the Brillouin zone can still be clearly assigned to vibrations of 1D $[Cl1_2]^-_\infty$ chains. However, when X is substituted with heavier halogens (Br and I), a spectral separation emerges in the PhDOS: the vibrational modes of Na and X are largely non-overlapping (Fig. S17b, c). This decoupling becomes more pronounced with increasing halogen mass, indicating that the 1D chains of Br and I in corresponding compounds are not only electronically isolated, but also vibrationally decoupled from the surrounding structure. The reduced hybridization of phonon modes can be attributed to both the mass contrast and the weaker bonding character between Na and heavier halogens, which together suppress inter-sublattice vibrational interactions. The observed vibrational decoupling is an additional strong factor contributing to the evidence of Peierls physics in the studied materials.

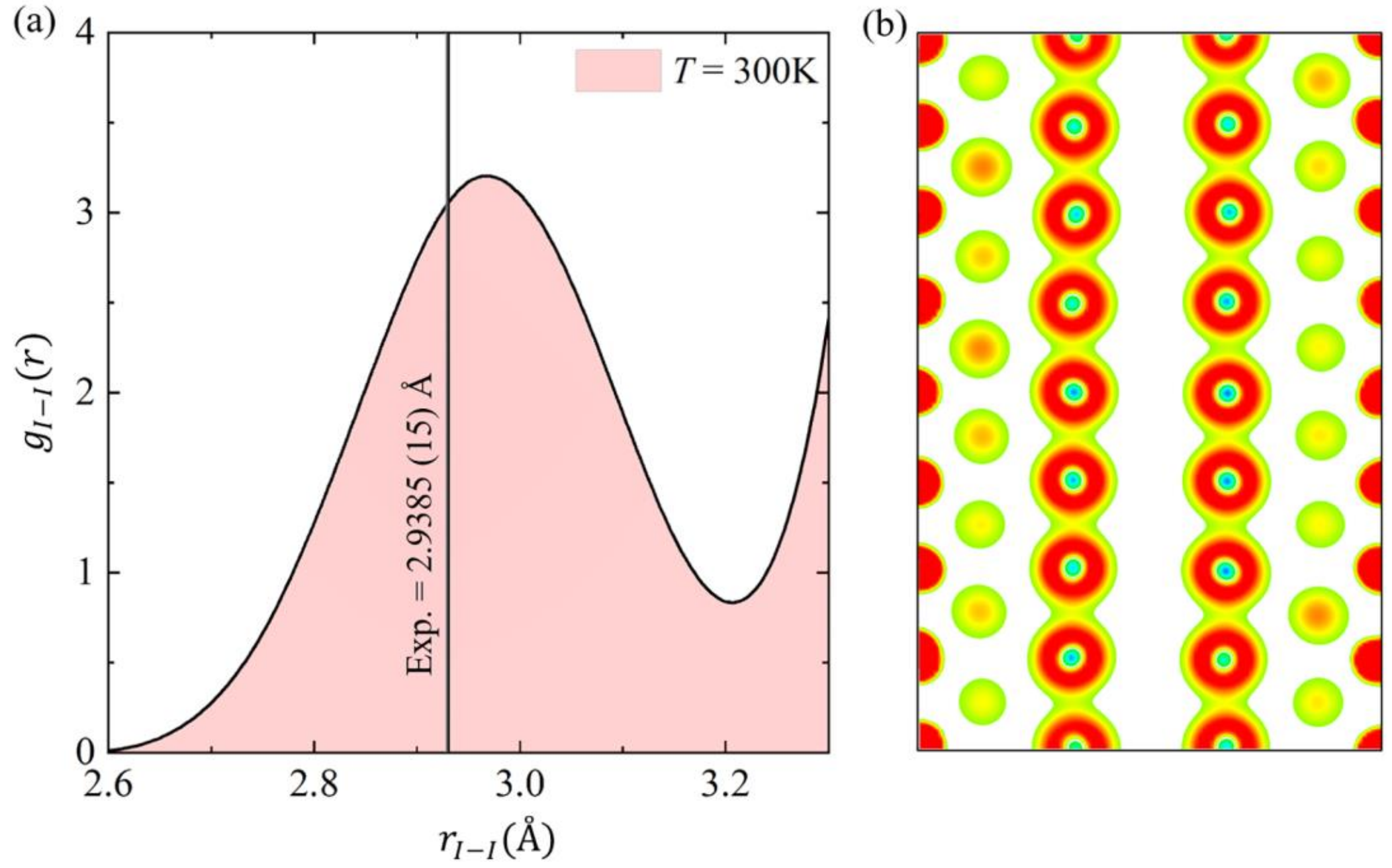

**Figure 3. Crystal structure and bonding of *hP*18-$Na_4I_5$, at temperature *T* = 300 K**. (a) Radial distribution functions (RDF) for *hP*18-$Na_4I_5$ at the pressure of 20 GPa and temperature of 300 K, obtained from AIMD simulations over a 50 ps trajectory. The vertical solid line at 2.9385 Å indicates the experimental I–I bond distances along the 1D iodine chain. (b) Averaged charge density from 10 uncorrelated thermalized AIMD snapshots at 300 K. The averaged charge density isosurfaces are visualized along the (110) plane, which intersects through the 1D iodine chains. The average charge on I-atoms on the 1D chain is -0.48 $e^-$/atom, derived from the Bader charge analysis of 10 AIMD snapshots at 300 K.

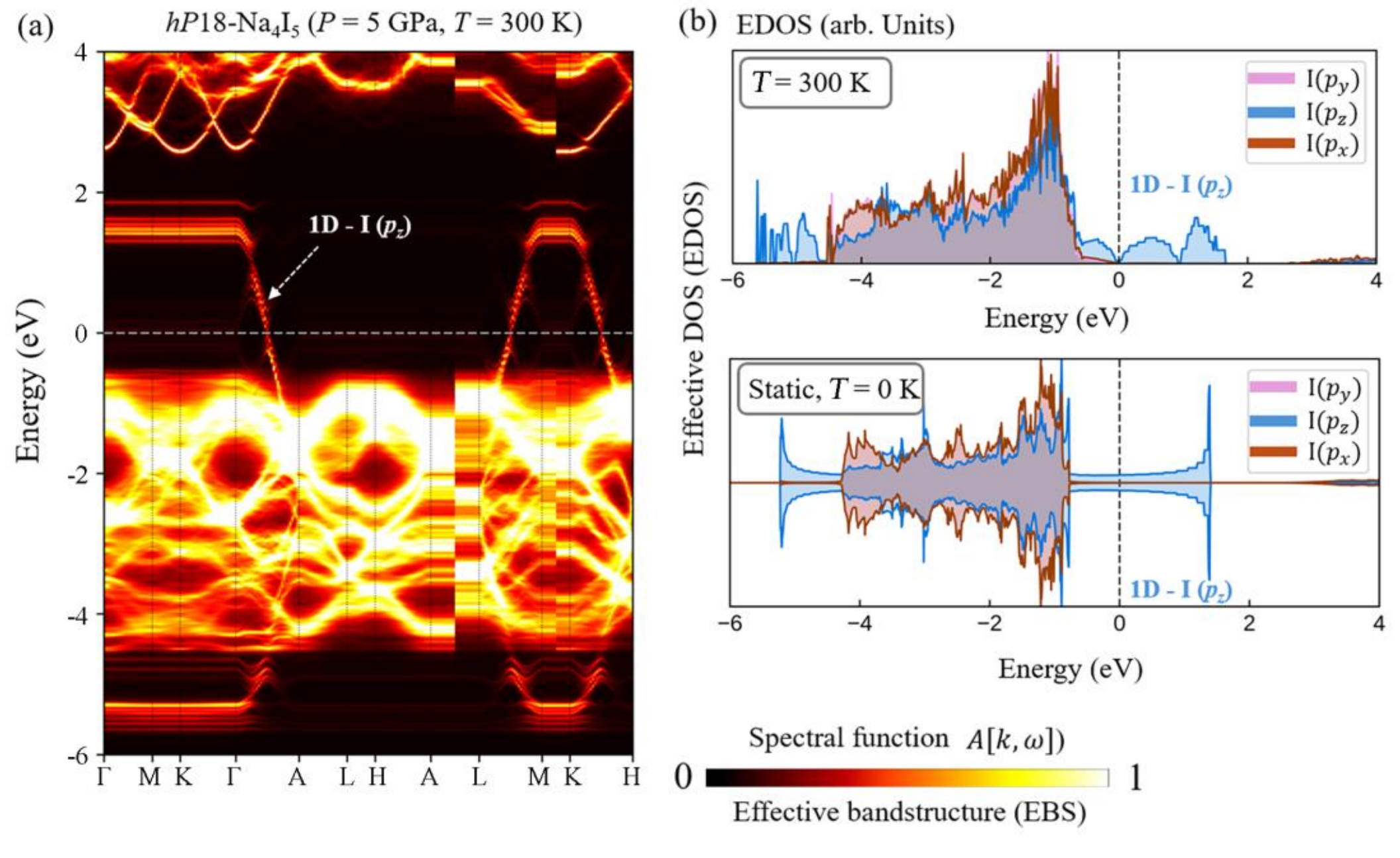

**Figure 4. Electronic structure of *hP*18-$Na_4I_5$ at temperature *T* = 300 K.** (a) Effective band structure (EBS) of *hP*18-$Na_4I_5$ at *T* = 300 K, and *P* = 5 GPa. The EBS was obtained by thermal averaging over 10 uncorrelated AIMD configurations at *T* = 300 K. The color scale represents the weight of the spectral function (ranging from 0 to 1), indicating the probability

of an electron at wavevector $k$ with energy $\omega$. (b) Corresponding effective density of states (EDOS) at $T = 300$ K, along with the DOS of the static lattice at $T = 0$ K. The projected density of states onto iodine 5$p$-orbitals reveals that the one-dimensional electronic states near the Fermi level originate primarily from the overlapping of I (5$p_z$) orbitals along the linear iodine chains. The opening of a pseudo gap at the Fermi energy is a result of strong electron-phonon coupling at finite temperature along the linear chain.

Let us discuss the influence of the finite temperature effects on the crystal structure and bonding of *hP*18-$Na_4X_5$, considering $Na_4I_5$ as a representative example. At room temperature, there are fluctuations of interatomic bond distances along the halogens chains as seen in the radial distribution function (RDF) calculated for *hP*18-$Na_4I_5$ (Fig. 3a). The RDF spread between I-I atoms along the chain at room temperature ($T = 300$ K) as compared to the static ($T = 0$ K) lattice leads to the electron density fluctuations around the average equilibrium positions of iodine atoms in the 1D $[I1_2]^-_\infty$ chain. However, the average Bader charges on each I1 atom in the chain remain almost the same – nearly 0.5 electron at 300 K (Fig. 3b). Remarkably, neither the interatomic bond distances along the $[I1_2]^-_\infty$ chains at finite temperatures, nor the nature of chemical bonding (particularly the shielding of the chains by the $Na^+$ ions and remaining $I^-$ ions), do not show any significant variation from the picture discussed for the static structure at 0 K (Fig. 2). On the contrary, the electronic structure of *hP*18-$Na_4I_5$ becomes qualitatively different (Fig. 4). In Fig. 4a, we show the effective band structure (EBS) of the compound calculated at $P = 5$ GPa and at $T = 300$ K. EBS appears to preserve most of the 1D band picture of *hP*18-$Na_4I_5$ arising from the I1 5$p_z$ orbitals in the $[I1_2]^-_\infty$ chains (see also Fig. S18). Fig. 4b shows the corresponding effective electronic DOS (EDOS) calculated at $T = 300$ K in comparison with the EDOS calculated for the static ($T = 0$ K) structure of *hP*18-$Na_4I_5$. At finite temperature one can clearly see the semi-metallic feature of the 1D bands at the Fermi energy (arising from the bands along the Γ-A direction in the EBS). In other words, geometrically ideal infinite 1D chains of equidistant atoms $[X_2]^-_\infty$ have semi-metallic electronic structure in *hP*18 $Na_4X_5$ compounds at finite temperatures.

## Conclusions

To conclude, we have suggested a path toward realizing one-dimensional (1D) electron quantum systems confined within a three-dimensional (3D) crystal structure. The well-known arguments of Landau and Peierls showed that linear chains of atoms are unstable with respect to phonon fluctuations even at zero temperature. A straightforward solution—to stabilize the chains by using a substrate or a cage—would affect their electronic properties and destroy their ideal one-dimensional nature. Surprisingly, we have found 3D structures that host linear chains of atoms without destroying the 1D character of their electronic structure. This is achieved through electronic isolation of the chains via a charge transfer mechanism. In the *hP*18-$Na_4X_5$ (X = I, Br, Cl) compounds, the (semi-)metallic 1D polyhalogen chains are shielded by an insulating network formed by closed-shell $Na^+$ and halogen $X^-$ ions, effectively creating atomically thin electric cables. The compounds we synthesized and studied show clear signatures of the Peierls effect (manifested in the formation of *i-hP*18 and *hP*36 $Na_4X_5$ phases), one of the main hallmarks of a 1D electronic structure. The natural presence of ordered arrays of 1D chains in these compounds provides a platform for extending the Luttinger liquid model toward higher dimensions.(*27*) The proposed approach should encourage the search for other systems with similar properties, opening a window into the world of truly 1D materials—much as the isolation of graphene opened the 2D world.

**Acknowledgments:** The authors acknowledge the Deutsches Elektronen-Synchrotron (DESY, PETRA III) and the European Synchrotron Radiation Facility (ESRF) for the provision of beamtime at the P02.2, and ID11 and ID15b beamlines, respectively. The authors acknowledge the ESRF for provision of beamtime at the ID15b beamtime for low temperature SCXRD measurements using Cryostat (proposal ID BLC-15329). The computations were enabled by resources provided by the National Academic Infrastructure for Supercomputing in Sweden (NAISS), partially funded by the Swedish Research Council under grant 2022-06725. We gratefully acknowledge the support of the staff at the National Supercomputing Centre (NSC) in Linköping and the staff at the Center for High Performance Computing (PDC) in Stockholm.

**Funding:**

Knut and Alice Wallenberg Foundation (Wallenberg Scholar grants no. KAW-2018.0194 and KAW-2023.0309) (Y.Y., N.J., Fe.T., Fl.T. and I.A.A)

Swedish Research Council (VR) Grant No. 2023-05358 (I.A.A)

Swedish Government Strategic Research Areas in Materials Science on Functional Materials at Linkoping University (Faculty Grant SFO-Mat-LiU No. 2009 00971) (N.D., I.A.A., Fl.T. and F.K.)

Swedish e-Science Center (SeRC) (N.D., I.A.A., Fl.T. and F.K.)

Deutsche Forschungsgemeinschaft (DFG; project DU 945–15/1) (N.D. and L.D.)

Swedish Research Council (VR) under grant 2023-05247 (S.I.S.)

European Research Council (ERC) under the Synergy Grant FASTCORR (project 854843). (S.I.S.)

Swedish Research Council (VR) program 2020-04630 (F.K.)

**Author contributions:** L.D., I.A.A., and N.D. designed the study; Y.Y., L.D., G.G., Al.A., An.A., X.L., A.A., F.I.A., K.G., H.F., J.B. conducted the experiments. Y.Y., G.G. performed low-temperature experiments on Na4I5 using a cryostat. L.D., G.G., X.L. performed laser heating on all samples. Theoretical calculations and analysis were performed by I.A.A., N.J., F.K., Fe.T., Y.Y., Fl.T., S.S., M.I.K. I.A.A.; L.D. and N.D. wrote the paper with contributions from all authors. All authors have given approval to the final version of the manuscript.

**Competing interests:** Authors declare that they have no competing interests.

**Data and materials availability:** Deposition Numbers 2451586-2451589, 2451900, 2451927, 2451928, 2451954, 2451962, 2451963, 2451971, and 2451972 contain the supplementary crystallographic data for this paper. These data can be obtained free of charge via the joint Cambridge Crystallographic Data Center (CCDC) and Fachinformationszentrum Karlsruhe.

**Supplementary Materials**

Supplementary Methods

Tables S1 to S11

Figs. S1 to S19

Supplementary Discussion 1 and 2

References (*28*–*53*)

# Supplementary Materials

## Supplementary Methods

Sample preparation

Two membrane-type diamond anvil cells (DACs) (*28*) equipped with 500 (DAC #1) and 350 μm (DAC #2) culet diamond anvils, and nine BX90-type screw-driven DACs (*29*) equipped with 250 (DAC #4-#6, #8-#11) and 120 μm (DAC #3, #7) culet diamond anvils were used. The sample chambers were formed by pre-indenting rhenium gaskets to ~20-40 μm thickness, and laser-drilling a hole of 60-240 μm, depending on the culet size. DAC #1 and #2 were loaded with same starting materials of NaI + $CI_4$ in the glove bag, and DAC #2 was further used for low temperature SCXRD measurements using Cryostat.(*30*) For experiments carried out for $Na_4Br_5$ and $Na_4Cl_5$ (DACs #3-#11), detailed loading can be found in Table S1. Double-sided laser-heating of the samples up to ~2300 K was performed for all DACs at the laboratory at the Bayerisches Geoinstitut (*31*), or at the ID15b beamline (*30*) of the European Synchrotron Radiation Facility (ESRF) with iodine (DAC #1 and #2), bromine (DAC #3-#6), carbon (DAC #10), or a piece of metal (DAC #7-#9, #11) employed as the laser light absorber. The pressure was measured *in situ* using the equation of state of rhenium (*32*) and the equation of state of B1-NaI.(*33*) Detailed information on experimental conditions and the observed $Na_4X_5$ phases is summarized in Table S1.

X-ray diffraction

Synchrotron X-ray diffraction measurements of the compressed samples were performed at ID11 ($\lambda$ = 0.28457 Å, beam size ~ 0.75 × 0.75 $\mu m^2$) and ID15b ($\lambda$ = 0.4099 Å, beam size ~ 2 × 2 $\mu m^2$) of the EBS-ESRF, and the P02.2 beamline ($\lambda$ = 0.2910 Å, beam size ~2.0 × 2.0 $\mu m^2$) of PETRA III. To determine the sample position for single-crystal X-ray diffraction data acquisition, a full X-ray diffraction mapping of the pressure chamber was performed. The sample positions displaying the greatest number of single-crystal reflections belonging to the phases of interest were chosen, and step-scans of 0.5° from −36° to +36° ω were performed. The CrysAlis$^{Pro}$ software (*34*) was utilized for the single-crystal data analysis. To calibrate the instrumental model in the CrysAlis$^{Pro}$ software, *i.e.* the sample-to-detector distance, detector's origin, offsets of the goniometer angles, and rotation of both the X-ray beam and detector around the instrument axis, we used a single crystal of vanadinite [$Pb_5(VO_4)_3Cl$, *P6₃/m* space group, *a* = 10.3174(30) Å, and *c* = 7.3378(30) Å] at ID15b, and orthoenstatite [$(Mg_{1.93}Fe_{0.06})(Si_{1.93}Al_{0.06})O_6$, *Pbca* space group, *a* = 8.8117(2) Å, *b* = 5.1832(10) Å, and *c* = 18.2391(3) Å] at the other beamlines. The DAFi program (*35*) was used for the search of reflections' groups belonging to individual single-crystal domains. The crystal structures were then solved and refined using the OLEX2 (*36*) software. The crystallite sizes were estimated from X-ray maps using the XDI software.(*37*) The crystallographic information is available in Tables S2-S7.

Density functional theory calculations

First-principles calculations were performed within the framework of density functional theory (DFT) as implemented in the Vienna Ab initio Simulation Package (VASP) (*38*) using the Projector-Augmented-Wave (PAW) method (*39, 40*), and the all-electron, full-potential code FHI-aims using the linear combination of numeric atom-centered orbitals (NAOs) method.(*41-43*) In our VASP calculations, the exchange-correlation effects were approximated using the Generalized Gradient Approximation (GGA) as proposed by Perdew–Burke–Ernzerhof (PBE) (*44*), an energy cut off was set to of 600 eV and a gamma centred 9x9x9 k-mesh was used for the structure relaxations until the atomic forces are < 1 meV/Å and residual stress < 0.1 GPa. The ground state charge density and electronic DOS were obtained using a denser 15x15x16 gamma centred k-mesh. The band structure of static lattices was evaluated from the ground state charge density along the high-symmetry band paths of the Brillouin zone of a hexagonal lattice.

While *hP*18 $Na_4X_5$ phases are metallic, *hP*36 $Na_4X_5$ phases have band gaps. Therefore, in addition to the GGA calculations, we carried out full structure relaxations including lattice parameters and atomic positions optimization using the hybrid DFT with the Heyd-Scuseria-Ernzerhof (HSE06) functional as implemented in FHI-aims.(*45*) For $Na_4Cl_5$ and $Na_4I_5$ compounds HSE calculations with typical default value of the mixing fraction of exact Hartree-Fock exchange with the semi-local PBE exchange functional $\alpha = 0.25$ were found to be sufficient for description of the cell volumes (Fig. S5c) and the X-X distances (d1, d2, d3) along the 1D halogen chains (Table S8). In the case of $Na_4Br_5$ compounds, we increased the parameter $\alpha$ to 0.375 and found that this slightly improves the lattice parameters and Br-Br distances (except for d3, which becomes marginally longer; Table S8).

For calculations using FHI-aims, the default suggestions for Light Atomic Species were employed, as they produced results comparable to those obtained with VASP for *hP*18 phases. Although PBE calculations reproduced the experimental lattice parameters of the *hP*36 $Na_4X_5$ phases, they resulted in large discrepancies of the X1-X1 distances in the linear halogen chains as compared to their experimental values, while hybrid DFT calculations reproduced them much better (see Table S8). Therefore, we report hybrid DFT HSE06 results for *hP*36 $Na_4X_5$ compounds. On the contrary, the crystal structure parameters for metallic *hP*18 $Na_4X_5$ compounds, primarily reported in the manuscript from hybrid DFT HSE06 calculations, are well reproduced already in the PBE GGA calculations. For long *ab initio* molecular dynamics (AIMD) simulations employed to study effects of temperature for *hP*18 phases, the use of hybrid DFT HSE06 functional becomes impractical. Comparing the average electronic density of states and the forces calculated with PBE GGA and HSE06 hybrid DFT for *hP*18-$Na_4I_5$ at 5 GPa (Fig. S19) we conclude that the former approximation is sufficient to adequately describe *hP*18 $Na_4X_5$ phases in AIMD simulations, as well as for calculations of their electronic band structure and electronic density of states, and the phonon dispersion relations. Note that at pressures ≤ 10 GPa, the relaxed geometries no longer retain the reference symmetries – $P6_3/mcm$ for *hP*18, and $P\bar{6}2c$ for *hP*36 (Table S11).

The atomic charges in Table S10 were calculated by applying the Bader's method as suggested in Ref (*46*). The crystal structure, charge density isosurfaces, and the ELF visualisations were made with the VESTA software.(*47*) The Fermi surface and the electronic band Wannierization in Fig. 2c and Fig. S13 have been performed using the full-potential local-orbital (FPLO) code version FPLO22.02-64.(*48, 49*) The numerical parameters were chosen to reproduce the electronic band structure obtained by VASP. Lattice dynamics calculations within the harmonic approximation (Fig. S9, Fig. S15) were performed using the finite displacement method in PHONOPY package (*50*), with force constants derived from FHI-aims or VASP DFT calculations. The phonon dispersion relations at finite temperature were obtained via the temperature-dependent effective potential (TDEP) method (*51, 52*) (Fig. S7, Fig. S17), using thermal configurations sampled from *ab initio* molecular dynamics (AIMD) simulations at the respective pressure-temperatures. All AIMD simulations were accelerated using VASP on-the-fly machine-learned force field (MLFF) method in a 2×2×4 supercell (288 atoms). The AIMD simulations were conducted at different pressure-temperature conditions for a particular phase of *hP*18-$Na_4X_5$ (X = I, Br, Cl) using a canonical ensemble with a Nosé–Hoover thermostat in a time step of 1 fs, and a total sampling period of 50 ps in a Γ-centered 2×2×2 k-point mesh. The final MLFF trained potential was used in a 3×3×4 (648 atoms) supercell to obtain the phonon dispersion relations for *hP*18-$Na_4X_5$ compounds (Fig. S7, Fig. S17). To calculate the effective band structure (EBS) and EDOS, 10 uncorrelated thermalized configurations from the AIMD simulations were used at room temperature (300 K). The supercell band structures were unfolded into the primitive Brillouin zone (BZ) of the static lattice at 0 K, using a modified version of the *easyunfold* Python code.(*53*)

**Table S1.** Summary of the performed high-pressure high-temperature experimental syntheses on $Na_4X_5$ (X = I, Br, Cl) compounds in laser-heated diamond anvil cells.

**$Na_4I_5$**

| DAC number | Culet diameter (μm) | Starting materials | Pressure | Temperature; ***in situ*** (K, ±5) | $Na_4I_5$ phase observed |
|---|---|---|---|---|---|
| #1 | 500 | NaI + $CI_4$ | 20(2)* | 298 | *hP*18 |
| | | | 12(2) | | *hP*18 |
| | | | 6(2) | | (PXRD) *hP*18** |
| | | | (Recovered)*** 2(2) | | (PXRD) *hP*18** |
| #2 | 350 | NaI + $CI_4$ | 24(2)* | 298 | *hP*18 |
| | | | 24(2) | 20 | *hP*18 |
| | | | 21(2) | 20 | *hP*18 |
| | | | 13(1) | 50 | *hP*18 |
| | | | 10(2) | 130 | *hP*18 |
| | | | (Recovered)*** 3(2) | 298 | (PXRD) *hP*18** |
| | | | 3(2)* | 298 | *hP*18 & *i-hP*18 |

**$Na_4Br_5$**

| DAC number | Culet diameter (μm) | Starting materials | Pressure | Temperature; ***in situ*** (K, ±5) | $Na_4Br_5$ phase observed |
|---|---|---|---|---|---|
| #3 | 120 | NaBr + $CBr_4$ | 73(1)* | 298 | *hP*18 |
| #4 | 250 | NaBr + $CBr_4$ | 48(1)* | 298 | *hP*18 & *hP*36 |
| #5 | 250 | NaBr + $CBr_4$ | 28(4)* | 298 | *hP*18 & *hP*36 |
| | | | 18(4) | | *hP*18 & *hP*36 |
| | | | 14(3) | | *hP*18 & *hP*36 & *i-hP*18 |
| | | | (Recovered)*** 6(4) | | (PXRD) *hP*18** |
| #6 | 250 | Au + NaBr + $CBr_4$ | 12(3)* | 298 | *hP*18 & *i-hP*18 |

**$Na_4Cl_5$**

| DAC number | Culet diameter (μm) | Starting materials | Pressure | Temperature; ***in situ*** (K, ±5) | $Na_4Cl_5$ phase observed |
|---|---|---|---|---|---|
| #7 | 120 | | 110(2)* | 298 | *hP*18 |
| | | Os + NaCl + $CCl_4$ | 109(2)* | | *hP*18 |
| | | | 92(6)* | | *hP*18 & *hP*36 |
| | | | 60(4)* | | *hP*18 & *hP*36 |
| #8 | 250 | Pt + NaCl + $CCl_4$ | 75(4)* | 298 | *hP*18 & *hP*36 |
| | | | 60(3)* | | *hP*18 & *hP*36 |
| #9 | 250 | Au + NaCl + $C_6Cl_6$ | 66(3)* | 298 | *hP*18 & *hP*36 |
| | | | 60(3)* | | *hP*18 & *hP*36 |
| | | | 47(3)* | | *hP*18 & *hP*36 & *i-hP*18 |
| #10 | 250 | C + NaCl + $CCl_4$ | 56(1)* | 298 | *hP*18 & *hP*36 |
| | | | 50(1)* | | *hP*18 & *hP*36 |
| | | | 42(1)* | | *hP*18 & *hP*36 & *i-hP*18 |
| #11 | 250 | Os + NaCl + $CCl_4$ | 36(5)* | 298 | *hP*18 & *i-hP*18 |
| | | | 30(2) | | *i-hP*18 |
| | | | 26(2) | | *i-hP*18 |
| | | | 24(2) | | *i-hP*18 |
| | | | 14(2) | | *i-hP*18 |
| | | | (Recovered)*** 2(2) | | - |

*At pressure points marked with the star*, laser heating (up to 2300(500) K) was performed prior SCXRD measurements to initiate chemical reactions or to facilitate crystal recrystallization.

**At pressure points marked with two stars**, the data quality was insufficient for a single-crystal refinement, and the phase identification based on powder PXRD (Le Bail fit was applied).

***"(Recovered)" indicates that before the XRD measurements at the specified pressure, the sample was fully decompressed to ambient pressure at room temperature by opening the DAC in an inert atmosphere (glove bag or glove box). The reported pressure corresponds to the value obtained after subsequently re-closing the DAC in the same bag/box following recovery.

**Table S2.** Crystal structure, data collection and refinement details of *hP*18-$Na_4I_5$ at selected pressure and temperature.

| **CSD Number** | 2451589 | 2451588 | 2451586 | 2451587 |
|---|---|---|---|---|
| **Crystal data** | | | | |
| Chemical formula | $Na_4I_5$ | $Na_4I_5$ | $Na_4I_5$ | $Na_4I_5$ |
| $M_r$ | 726.46 | 726.46 | 726.46 | 726.46 |
| Crystal system, space group | Hexagonal, $P6_3/mcm$ | Hexagonal, $P6_3/mcm$ | Hexagonal, $P6_3/mcm$ | Hexagonal, $P6_3/mcm$ |
| Temperature (K) | 298 | 50 | 130 | 298 |
| Pressure (GPa) | 24(2) | 13(1) | 10(2) | 3(2) |
| *a, c* (Å) | 8.490(4), 5.7769(9) | 8.7985(13), 5.9641(7) | 8.909(5), 6.060(3) | 10.149(4), 6.581(5) |
| $V$ (Å$^3$) | 360.6(3) | 399.85(13) | 416.5(5) | 587.1(7) |
| $Z$ | 2 | 2 | 2 | 2 |
| Radiation type | Synchrotron, λ = 0.4099 Å | | | |
| Crystal size (mm) | 0.004×0.004×0.004 | 0.003×0.003×0.003 | 0.003×0.003×0.003 | 0.003×0.003×0.003 |
| **Data collection** | | | | |
| Diffractometer | ESRF ID15b, EIGER2 X 9M CdTe detector | | | |
| No. of measured, independent and observed [$I > 2\sigma(I)$] reflections | 621, 175, 153 | 451, 216, 144 | 306, 175, 121 | 741, 314, 169 |
| $R_{int}$ | 0.027 | 0.081 | 0.047 | 0.065 |
| $(\sin \theta/\lambda)_{max}$ (Å$^{-1}$) | 0.724 | 0.716 | 0.787 | 0.827 |
| **Refinement** | | | | |
| $R[F^2 > 2\sigma(F^2)]$, $wR(F^2)$, $S$ | 0.028, 0.063, 1.15 | 0.071, 0.189, 1.06 | 0.059, 0.181, 1.08 | 0.073, 0.210, 0.92 |
| No. of reflections | 175 | 216 | 175 | 314 |
| No. of parameters | 13 | 13 | 13 | 13 |
| $\Delta\rho_{max}$, $\Delta\rho_{min}$ (e Å$^{-3}$) | 1.02, -0.98 | 2.61, -2.60 | 2.81, -2.53 | 2.02, -1.48 |
| **Crystal Structure** | | | | |
| Wyckoff Site | Na1: 2*b*<br>Na2: 6*g*<br>I1: 4*d*<br>I2: 6*g* | Na1: 2*b*<br>Na2: 6*g*<br>I1: 4*d*<br>I2: 6*g* | Na1: 2*b*<br>Na2: 6*g*<br>I1: 4*d*<br>I2: 6*g* | Na1: 2*b*<br>Na2: 6*g*<br>I1: 4*d*<br>I2: 6*g* |
| Fractional atomic coordinates (x y z) | Na1: 0 0 0<br>Na2: 0.6185(8) 0 1/4<br>I1: 1/3 2/3 0<br>I2: 0.27244(12) 0 1/4 | Na1: 0 0 0<br>Na2: 0.6191(18) 0 1/4<br>I1: 1/3 2/3 0<br>I2: 0.2711(3) 0 1/4 | Na1: 0 0 0<br>Na2: 0.6198(16) 0 1/4<br>I1: 1/3 2/3 0<br>I2: 0.2711(3) 0 1/4 | Na1: 0 0 0<br>Na2: 0.612(3) 0 1/4<br>I1: 1/3 2/3 0<br>I2: 0.25805(16) 0 1/4 |
| $U_{iso}$ (Å$^2$) | Na1: 0.016(2)<br>Na2: 0.0220(14)<br>I1: 0.0173(3)<br>I2: 0.0177(3) | Na1: 0.019(5)<br>Na2: 0.024(3)<br>I1: 0.0229(9)<br>I2: 0.0202(8) | Na1: 0.026(5)<br>Na2: 0.022(3)<br>I1: 0.0236(9)<br>I2: 0.0206(8) | Na1: 0.061(8)<br>Na2: 0.154(13)<br>I1: 0.0597(10)<br>I2: 0.0414(7) |

**Table S3.** Crystal structure, data collection and refinement details of *i*-*hP*18-$Na_4I_5$ at 3(2) GPa.

| | |
|---|---|
| **CSD Number** | 2451900 |
| **Crystal data** | |
| Chemical formula | $Na_4I_5$ |
| $M_r$ | 726.46 |
| Crystal system, space group | Hexagonal, $P6_3/mcm(00\gamma)00ss$ |
| Temperature (K) | 298 |
| Pressure (GPa) | 3(2) |
| Wave vectors | $\mathbf{q} = 0.4858(8)\mathbf{c}^*$ |
| $a$, $c$ (Å) | 10.111(2), 6.5847(5) |
| $V$ (Å$^3$) | 582.92(18) |
| $Z$ | 2 |
| Radiation type | Synchrotron, $\lambda$ = 0.4099 Å |
| Crystal size (mm) | 0.003 × 0.003 × 0.003 |
| **Data collection** | |
| Diffractometer | ESRF ID15b, EIGER2 X 9M CdTe detector |
| No. of main reflections: measured, independent and observed [$I > 3\sigma(I)$] | 1159, 392, 366 |
| No. of 1$^{st}$-order satellite reflections: measured, independent and observed [$I > 3\sigma(I)$] | 1669, 565, 392 |
| $R_{int}$ | 0.050 |
| $(\sin \theta/\lambda)_{max}$ (Å$^{-1}$) | 0.830 |
| **Refinement** | |
| $R(obs)_{main + satellites}$ / $wR(all)_{main + satellites}$ | 0.070 / 0.098 |
| $R(obs)_{main}$ / $wR(all)_{main}$ | 0.060 / 0.095 |
| $R(obs)_{1st\ order\ satellites}$ / $wR(all)_{1st\ order\ satellites}$ | 0.100 / 0.104 |
| No. of reflections | 957 |
| No. of parameters | 18 |
| $\Delta\rho_{max}$, $\Delta\rho_{min}$ (e Å$^{-3}$) | 4.61, -3.90 |
| **Crystal Structure** | |
| Na1 ($x$ $y$ $z$) | (0 0 0) |
| Na2 ($x$ $y$ $z$)<br>$A_1(x)$<br>$A_2(y)$ | (0.6109(8) 0 1/4)<br>0.017(2)<br>0.0087(10) |
| I1 ($x$ $y$ $z$)<br>$A_1(z)$ | (1/3 2/3 0)<br>0.0443(3) |
| I2 ($x$ $y$ $z$)<br>$A_1(x)$<br>$A_2(y)$ | (0.25778(7) 0 1/4)<br>0.00282(5)<br>0.00563(10) |

**Table S4.** Crystal structure, data collection and refinement details of *hP*36-$Na_4Br_5$ at selected pressures.

| **CSD Number** | 2451928 | 2451927 |
|---|---|---|
| **Crystal data** | | |
| Chemical formula | $Na_4Br_5$ | $Na_4Br_5$ |
| $M_r$ | 491.51 | 491.51 |
| Crystal system, space group | Hexagonal, $P\bar{6}2c$ | Hexagonal, $P\bar{6}2c$ |
| Temperature (K) | 298 | 298 |
| Pressure (GPa) | 28(4) | 14(3) |
| *a, c* (Å) | 7.8357(7), 10.4204(14) | 8.1983(13), 10.884(2) |
| $V$ (Å$^3$) | 554.08(12) | 633.6(2) |
| $Z$ | 4 | 4 |
| Radiation type | Synchrotron, λ = 0.4099 Å | |
| Crystal size (mm) | 0.003×0.003×0.003 | 0.003×0.003×0.003 |
| **Data collection** | | |
| Diffractometer | ESRF ID15b, EIGER2 X 9M CdTe detector | |
| No. of measured, independent and observed [$I > 2\sigma(I)$] reflections | 1528, 665, 535 | 1485, 751, 573 |
| $R_{int}$ | 0.018 | 0.018 |
| $(\sin \theta/\lambda)_{max}$ (Å$^{-1}$) | 0.898 | 0.833 |
| **Refinement** | | |
| $R[F^2 > 2\sigma(F^2)]$, $wR(F^2)$, $S$ | 0.022, 0.058, 1.07 | 0.044, 0.142, 1.07 |
| No. of reflections | 665 | 751 |
| No. of parameters | 32 | 32 |
| $\Delta\rho_{max}$, $\Delta\rho_{min}$ (e Å$^{-3}$) | 1.56, −0.88 | 1.81, −1.60 |
| **Crystal Structure** | | |
| Wyckoff Site | Na1: 4*e*<br>Na2a: 6*h*<br>Na2b: 6*g*<br>Br1a: 4*f*<br>Br1b: 4*f*<br>Br2a: 6*h*<br>Br2b: 6*g* | Na1: 4*e*<br>Na2a: 6*h*<br>Na2b: 6*g*<br>Br1a: 4*f*<br>Br1b: 4*f*<br>Br2a: 6*h*<br>Br2b: 6*g* |
| Fractional atomic coordinates (x y z) | Na1: 0 0 0.1201(8)<br>Na2a: 0.3857(6) 0.0084(2) 1/4<br>Na2b: 0.6176(6) 0 0<br>Br1a: 1/3 2/3 0.13192(13)<br>Br1b: 1/3 2/3 0.61719(13)<br>Br2a: 0.27444(6) 0.27542(16) 1/4<br>Br2b: 0.27500(16) 0 0 | Na1: 0 0 0.1328(7)<br>Na2a: 0.3892(9) 0.0148(7) 1/4<br>Na2b: 0.6200(7) 0 0<br>Br1a: 1/3 2/3 0.13751(12)<br>Br1b: 1/3 2/3 0.61485(13)<br>Br2a: 0.2725(2) 0.2741(2) 1/4<br>Br2b: 0.27308(19) 0 0 |
| $U_{iso}$ (Å$^2$) | Na1: 0.0102(6)<br>Na2a: 0.0107(11)<br>Na2b: 0.0142(12)<br>Br1a: 0.01062(15)<br>Br1b: 0.01001(15) | Na1: 0.0208(10)<br>Na2a: 0.0301(16)<br>Na2b: 0.0243(15)<br>Br1a: 0.0219(3)<br>Br1b: 0.0251(3) |
| | Br2a: 0.0103(3)<br>Br2b: 0.0097(3) | Br2a: 0.0207(4)<br>Br2b: 0.0230(4) |

**Table S5.** Crystal structure, data collection and refinement details of *i-hP*18-$Na_4Br_5$ at 14(3) GPa.

| **CSD Number** | 2451954 |
|---|---|
| **Crystal data** | |
| Chemical formula | $Na_4Br_5$ |
| $M_r$ | 491.51 |
| Crystal system, space group | Hexagonal, $P6_3/mcm(00\gamma)00ss$ |
| Temperature (K) | 298 |
| Pressure (GPa) | 14(3) |
| Wave vectors | **q** = 0.4780(19)**c*** |
| *a, c* (Å) | 8.2094(18), 5.4079(11) |
| $V$ (Å$^3$) | 315.63(12) |
| $Z$ | 2 |
| Radiation type | Synchrotron, λ = 0.4099 Å |
| Crystal size (mm) | 0.003 × 0.003 × 0.003 |
| **Data collection** | |
| Diffractometer | ESRF ID15b, EIGER2 X 9M CdTe detector |
| No. of main reflections: measured, independent and observed [$I > 3\sigma(I)$] | 460, 200, 189 |
| No. of 1$^{st}$-order satellite reflections: measured, independent and observed [$I > 3\sigma(I)$] | 800, 319, 137 |
| $R_{int}$ | 0.056 |
| $(\sin\theta/\lambda)_{max}$ (Å$^{-1}$) | 0.832 |
| **Refinement** | |
| $R(obs)_{main + satellites}$ / $wR(all)_{main + satellites}$ | 0.046 / 0.069 |
| $R(obs)_{main}$ / $wR(all)_{main}$ | 0.043 / 0.065 |
| $R(obs)_{1st\ order\ satellites}$ / $wR(all)_{1st\ order\ satellites}$ | 0.062 / 0.078 |
| No. of reflections | 519 |
| No. of parameters | 18 |
| $\Delta\rho_{max}$, $\Delta\rho_{min}$ (e Å$^{-3}$) | 2.31, −2.14 |
| **Crystal Structure** | |
| Na1 (*x y z*) | (0 0 0) |
| Na2 (*x y z*)<br>$A_1(x)$<br>$A_2(y)$ | (0.6192(4) 0 1/4)<br>0.0069(3)<br>-0.0069(3) |
| Br1 (*x y z*)<br>$A_1(z)$ | (1/3 2/3 0)<br>0.0282(2) |
| Br2 (*x y z*)<br>$A_1(x)$<br>$A_2(y)$ | (0.27350(8) 0 1/4)<br>0.00133(13)<br>0.00067(7) |

**Table S6.** Crystal structure, data collection and refinement details of *hP*36-$Na_4Cl_5$ at selected pressures.

| CSD Number | 2451962 | 2451963 |
|---|---|---|
| **Crystal data** | | |
| Chemical formula | $Na_4Cl_5$ | $Na_4Cl_5$ |
| $M_r$ | 269.21 | 269.21 |
| Crystal system, space group | Hexagonal, $P\bar{6}2c$ | Hexagonal, $P\bar{6}2c$ |
| Temperature (K) | 298 | 298 |
| Pressure (GPa) | 92(6) | 47(3) |
| *a, c* (Å) | 6.9672(14), 9.1339(9) | 7.3914(14), 9.6722(12) |
| $V$ (Å$^3$) | 383.98(16) | 457.62(18) |
| $Z$ | 4 | 4 |
| Radiation type | Synchrotron, λ = 0.28457 Å | Synchrotron, λ = 0.4099 Å |
| Crystal size (mm) | 0.003×0.003×0.003 | 0.003×0.003×0.003 |
| **Data collection** | | |
| Diffractometer | ESRF ID11, Dectris Eiger2 X CdTe 4M | ESRF ID15b, EIGER2 X 9M CdTe detector |
| No. of measured, independent and observed [$I > 2\sigma(I)$] reflections | 2087, 755, 477 | 788, 388, 274 |
| $R_{int}$ | 0.035 | 0.027 |
| $(\sin \theta/\lambda)_{max}$ (Å$^{-1}$) | 1.102 | 0.883 |
| **Refinement** | | |
| $R[F^2 > 2\sigma(F^2)]$, $wR(F^2)$, $S$ | 0.043, 0.111, 1.00 | 0.059, 0.194, 1.08 |
| No. of reflections | 755 | 388 |
| No. of parameters | 32 | 32 |
| $\Delta\rho_{max}$, $\Delta\rho_{min}$ (e Å$^{-3}$) | 0.89, −0.71 | 1.25, −1.01 |
| **Crystal Structure** | | |
| Wyckoff Site | Na1: 4*e*<br>Na2a: 6*h*<br>Na2b: 6*g*<br>Cl1a: 4*f*<br>Cl1b: 4*f*<br>Cl2a: 6*h*<br>Cl2b: 6*g* | Na1: 4*e*<br>Na2a: 6*h*<br>Na2b: 6*g*<br>Cl1a: 4*f*<br>Cl1b: 4*f*<br>Cl2a: 6*h*<br>Cl2b: 6*g* |
| Fractional atomic coordinates (x y z) | Na1: 0 0 0.1248(10)<br>Na2a: 0.3851(8) 0.0101(4) 1/4<br>Na2b: 0.6154(8) 0 0<br>Cl1a: 1/3 2/3 0.1349(3)<br>Cl1b: 1/3 2/3 0.6156(3)<br>Cl2a: 0.2785(5) 0.2801(3) 1/4<br>Cl2b: 0.2777(5) 0 0 | Na1: 0 0 0.1249(10)<br>Na2a: 0.3885(10) 0.0140(10) 1/4<br>Na2b: 0.6167(9) 0 0<br>Cl1a: 1/3 2/3 0.1353(4)<br>Cl1b: 1/3 2/3 0.6147(4)<br>Cl2a: 0.2764(5) 0.2781(3) 1/4<br>Cl2b: 0.2760(5) 0 0 |
| $U_{iso}$ (Å$^2$) | Na1: 0.0107(4)<br>Na2a: 0.0122(9)<br>Na2b: 0.0119(10)<br>Cl1a: 0.0111(3) | Na1: 0.0139(10)<br>Na2a: 0.0149(11)<br>Na2b: 0.0136(11)<br>Cl1a: 0.0155(7) |
| | Cl1b: 0.0109(3)<br>Cl2a: 0.0118(6)<br>Cl2b: 0.0099(5) | Cl1b: 0.0145(6)<br>Cl2a: 0.0111(9)<br>Cl2b: 0.0127(11) |

**Table S7.** Crystal structure, data collection and refinement details of *i*-*hP*18-$Na_4Cl_5$ at selected pressures.

| **CSD Number** | 2451972 | 2451971 |
|---|---|---|
| **Crystal data** | | |
| Chemical formula | $Na_4Cl_5$ | $Na_4Cl_5$ |
| $M_r$ | 269.21 | 269.21 |
| Crystal system, space group | Hexagonal, $P6_3/mcm(00\gamma)00ss$ | Hexagonal, $P6_3/mcm(00\gamma)00ss$ |
| Temperature (K) | 298 | 298 |
| Pressure (GPa) | 47(3) | 14(2) |
| Wave vectors | $\mathbf{q} = 0.481(3)\mathbf{c}^*$ | $\mathbf{q} = 0.454(4)\mathbf{c}^*$ |
| *a, c* (Å) | 7.3997(14), 4.8452(14) | 7.9731(18), 5.156(3) |
| $V$ (Å$^3$) | 229.76(9) | 283.86(19) |
| $Z$ | 2 | 2 |
| Radiation type | Synchrotron, λ = 0.4099 Å | |
| Crystal size (mm) | 0.003 × 0.003 × 0.003 | 0.003 × 0.003 × 0.003 |
| **Data collection** | | |
| Diffractometer | ESRF ID15b, EIGER2 X 9M CdTe detector | |
| No. of main reflections: measured, independent and observed [$I > 3\sigma(I)$] | 511, 154, 125 | 491, 114, 93 |
| No. of 1st-order satellite reflections: measured, independent and observed [$I > 3\sigma(I)$] | 912, 255, 74 | 909, 164, 49 |
| $R_{int}$ | 0.091 | 0.090 |
| $(\sin \theta/\lambda)_{max}$ (Å$^{-1}$) | 0.882 | 0.693 |
| **Refinement** | | |
| $R(obs)_{main + satellites}$ / $wR(all)_{main + satellites}$ | 0.072 / 0.077 | 0.057 / 0.066 |
| $R(obs)_{main}$ / $wR(all)_{main}$ | 0.072 / 0.076 | 0.047 / 0.056 |
| $R(obs)_{1st\ order\ satellites}$ / $wR(all)_{1st\ order\ satellites}$ | 0.067 / 0.080 | 0.122 / 0.133 |
| No. of reflections | 409 | 278 |
| No. of parameters | 18 | 18 |
| $\Delta\rho_{max}$, $\Delta\rho_{min}$ (e Å$^{-3}$) | 4.51, −2.59 | 1.19, −1.18 |
| **Crystal Structure** | | |
| Na1 (*x y z*) | (0 0 0) | (0 0 0) |
| Na2 (*x y z*)<br>$A_1(x)$<br>$A_2(y)$ | (0.6179(3) 0 1/4)<br>0.0079(3)<br>0.0158(6) | (0.6212(3) 0 1/4)<br>0.0231(6)<br>0.0116(3) |
| Cl1 (*x y z*)<br>$A_1(z)$ | (1/3 2/3 0)<br>0.0371(6) | (1/3 2/3 0)<br>0.0506(11) |
| Cl2 (*x y z*)<br>$A_1(x)$<br>$A_2(y)$ | (0.27622(17) 0 1/4)<br>0.00114(19)<br>0.00114(19) | (0.27399(18) 0 1/4)<br>0.00141(19)<br>0.0028(4) |

**Table S8.** Comparison between the calculated (PBE and HSE06) and experimental (Exp.) lattice parameters and bond lengths along the 1D halogen chains in *hP*18 and *hP*36-$Na_4X_5$ phases at selected pressures.

| ***hP*18-$Na_4Cl_5$** | | | | | | | |
|---|---|---|---|---|---|---|---|
| Pressure (GPa) | Method | $a$ (Å) | $c$ (Å) | $V$ (Å$^3$) | $d_1$ (Å) | | |
| 50(1) | Exp. | 7.329(2) | 4.7757(18) | 222.14(4) | 2.3879(9) | | |
| 50 | PBE | 7.275 | 4.748 | 217.609 | 2.374 | | |
| | HSE06 | 7.360 | 4.793 | 224.805 | 2.397 | | |
| ***hP*36-$Na_4Cl_5$** | | | | | | | |
| Pressure (GPa) | Method | $a$ (Å) | $c$ (Å) | $V$ (Å$^3$) | $d_1$ (Å) | $d_2$ (Å) | $d_3$ (Å) |
| 50(1) | Exp. | 7.351(3) | 9.581(4) | 448.3(4) | 2.137(6) | 2.422(5) | 2.595(6) |
| 50 | PBE | 7.369 | 9.604 | 451.707 | 2.279 | 2.400 | 2.526 |
| | HSE06 | 7.357 | 9.610 | 450.461 | 2.130 | 2.408 | 2.664 |
| 92(6) | Exp. | 6.9672(14) | 9.1339(9) | 383.98(16) | 2.103(6) | 2.288(4) | 2.455(6) |
| 90 | PBE | 7.006 | 9.142 | 388.580 | 2.283 | 2.285 | 2.288 |
| | HSE06 | 6.998 | 9.144 | 387.833 | 2.101 | 2.290 | 2.463 |
| ***hP*18-$Na_4Br_5$** | | | | | | | |
| Pressure (GPa) | Method | $a$ (Å) | $c$ (Å) | $V$ (Å$^3$) | $d_1$ (Å) | | |
| 48(1) | Exp. | 7.6520(14) | 5.1262(19) | 259.94(14) | 2.5631(10) | | |
| 50 | PBE | 7.682 | 5.107 | 260.990 | 2.554 | | |
| 50 | HSE06 | 7.673 | 5.094 | 259.725 | 2.547 | | |
| ***hP*36-$Na_4Br_5$** | | | | | | | |
| Pressure (GPa) | Method | $a$ (Å) | $c$ (Å) | $V$ (Å$^3$) | $d_1$ (Å) | $d_2$ (Å) | $d_3$ (Å) |
| 18(4) | Exp. | 7.9548(8) | 10.5815(18) | 579.88(15) | 2.438(4) | 2.663(3) | 2.817(4) |
| 20 | HSE06 $\alpha = 0.125$ | 8.251 | 10.908 | 643.143 | 2.536 | 2.719 | 2.933 |
| | HSE06 $\alpha = 0.25$ | 8.238 | 10.896 | 640.208 | 2.472 | 2.716 | 2.992 |
| | HSE06 | 8.223 | 10.884 | 637.226 | 2.422 | 2.715 | 3.030 |
| | α = 0.375 | | | | | | |
| | HSE06 α = 0.5 | 8.208 | 10.874 | 634.246 | 2.382 | 2.718 | 3.056 |
| 48(1) | Exp. | 7.6476(10) | 10.2561(12) | 519.47(15) | 2.434(4) | 2.563(3) | 2.696(4) |
| 50 | PBE | 7.648 | 10.239 | 518.661 | 2.560 | 2.560 | 2.560 |
| | HSE06 α = 0.375 | 7.671 | 10.210 | 520.097 | 2.370 | 2.551 | 2.735 |

| ***hP*18-$Na_4I_5$** | | | | | |
|---|---|---|---|---|---|
| Pressure (GPa) | Method | *a* (Å) | *c* (Å) | *V* (Å$^3$) | $d_1$ (Å) |
| 20(2) | Exp. | 8.636(3) | 5.877(3) | 379.6(3) | 2.9385(15) |
| 20 | PBE | 8.759 | 5.922 | 393.524 | 2.961 |
| 20 | HSE06 | 8.778 | 5.916 | 394.781 | 2.958 |

**Table S9.** The parameters of the third order Birch-Murnaghan equations of states (3BM EOSes) used to fit the theoretical HSE06 P-V data for $Na_4X_5$.

| Compound | Bulk modulus $K_0$ (GPa) | $K'$ | $V_0$ (Å$^3$/atom) |
|---|---|---|---|
| *hP*18-$Na_4Cl_5$ | 33.8(6) | 4.264(18) | 20.62(7) |
| *hP*36-$Na_4Cl_5$ | 27.5(8) | 4.43(3) | 21.45(10) |
| *hP*18-$Na_4Br_5$ | 21.7(15) | 4.63(9) | 25.8(3) |
| *hP*36-$Na_4Br_5$ | 17.9(17) | 4.79(11) | 26.8(4) |
| *hP*18-$Na_4I_5$ | 15.3(9) | 4.87(13) | 34.0(2) |
| *hP*36-$Na_4I_5$ | 12.1(3) | 5.03(10) | 35.40(7) |

**Table S10.** Calculated Bader charge for specified atoms in B2 NaCl, B1 and B33 NaBr, B1-NaI, *hP*18-$Na_4X_5$, and *hP*36-$Na_4X_5$ at different pressures using HSE06.

| Compound | Atom name | Wyckoff Site | Bader charge | |
|---|---|---|---|---|
| | | | 50 GPa | 90 GPa |
| B2-NaCl | Na1 | 1*a* | 0.83 | 0.81 |
| | Cl1 | 1*b* | -0.83 | -0.81 |
| *hP*18-$Na_4Cl_5$ | Na1 | 2*b* | 0.79 | 0.77 |
| | Na2 | 6*g* | 0.83 | 0.81 |
| | Cl1 | 4*d* | -0.42 | -0.41 |
| | Cl2 | 6*g* | -0.81 | -0.79 |
| *hP*36-$Na_4Cl_5$ | Na1 | 4*e* | 0.79 | 0.77 |
| | Na2a | 6*h* | 0.83 | 0.81 |
| | Na2b | 6*g* | 0.83 | 0.81 |
| | Cl1a | 4*f* | -0.23 | -0.31 |
| | Cl1b | 4*f* | -0.63 | -0.53 |
| | Cl2a | 6*h* | -0.81 | -0.79 |
| | Cl2b | 6*g* | -0.81 | -0.79 |

| Compound | Atom name | Wyckoff Site | Bader charge | |
|---|---|---|---|---|
| | | | 20 GPa | 40 GPa |
| B1-NaBr | Na1 | 4*a* | 0.81 | - |
| | Br1 | 4*b* | -0.81 | - |
| B33-NaBr | Na1 | 4*c* | - | 0.80 |
| | Br1 | 4*c* | - | -0.80 |
| *hP*18-$Na_4Br_5$ | Na1 | 2*b* | 0.80 | 0.79 |
| | Na2 | 6*g* | 0.85 | 0.83 |
| | Br1 | 4*d* | -0.43 | -0.42 |
| | Br2 | 6*g* | -0.83 | -0.81 |
| *hP*36-$Na_4Br_5$ | Na1 | 4*e* | 0.80 | 0.78 |
| | Na2a | 6*h* | 0.85 | 0.83 |
| | Na2b | 6*g* | 0.85 | 0.82 |
| | Br1a | 4*f* | -0.22 | -0.30 |
| | Br1b | 4*f* | -0.65 | -0.55 |
| | Br2a | 6*h* | -0.82 | -0.80 |
| | Br2b | 6*g* | -0.82 | -0.80 |

| Compound | Atom name | Wyckoff Site | Bader charge |
|---|---|---|---|
| | | | 10 GPa |
| B1-NaI | Na1 | 4*a* | 0.81 |
| | I1 | 4*b* | -0.81 |
| *hP*18-$Na_4I_5$ | Na1 | 2*b* | 0.80 |
| | Na2 | 6*g* | 0.85 |
| | I1 | 4*d* | -0.43 |
| | I2 | 6*g* | -0.83 |
| *hP*36-$Na_4I_5$ | Na1 | 4*e* | 0.80 |
| | Na2a | 6*h* | 0.85 |
| | Na2b | 6*g* | 0.85 |
| | I1a | 4*f* | -0.32 |
| | I1b | 4*f* | -0.56 |
| | I2a | 6*h* | -0.83 |
| | I2b | 6*g* | -0.83 |

**Table S11.** Summary of the DFT-relaxed symmetries using HSE06 of the $Na_4X_5$ compounds.

| Compound | Pressure (GPa) | HSE06 | |
|---|---|---|---|
| | | Symmetry tolerance | Space group |
| *hP*18-$Na_4Cl_5$ | 0 | 5e-03 | $P2_1/m$ |
| | 10-120 | 5e-03 | $P6_3/mcm$ |
| *hP*36-$Na_4Cl_5$ | 0 | 5e-02 | *Ama*2 |
| | 10 | 5e-02 | $P\bar{6}2c$ |
| | | 5e-03 | *Ama*2 |
| | 20-90 | 5e-03 | $P\bar{6}2c$ |
| *hP*18-$Na_4Br_5$ | 0 | 5e-03 | $P2_1/m$ |
| | 10 | 5e-02 | $P6_3/mcm$ |
| | | 5e-03 | *Cmcm* |
| | 20-80 | 5e-03 | $P6_3/mcm$ |
| *hP*36-$Na_4Br_5$ | 0 | 5e-02 | *Ama*2 |
| | 10 | 5e-02 | $P\bar{6}2c$ |
| | | 5e-03 | *Ama*2 |
| | 20-70 | 5e-03 | $P\bar{6}2c$ |
| | 80 | 5e-02 | $P\bar{6}2c$ |
| *hP*18-$Na_4I_5$ | 0 | 5e-03 | *Cmcm* |
| | 10-30 | 5e-02 | $P6_3/mcm$ |
| *hP*36-$Na_4I_5$ | 0 | 5e-02 | *Ama*2 |
| | 10 | 5e-02 | $P\bar{6}2c$ |

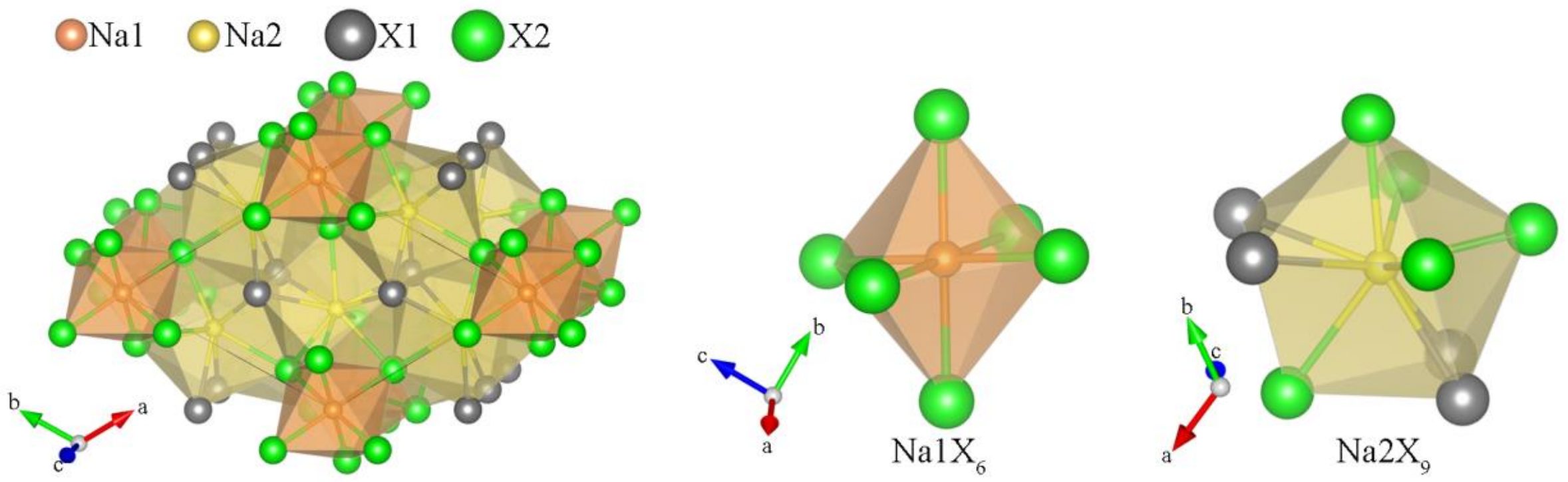

**Fig. S1.** Polyhedral model of *hP*18-$Na_4X_5$ (X = I, Br, Cl) and coordination polyhedra for Na1 and Na2 atoms. Na1 atoms are orange, Na2 atoms are yellow, X1 atoms are grey, X2 atoms are green.

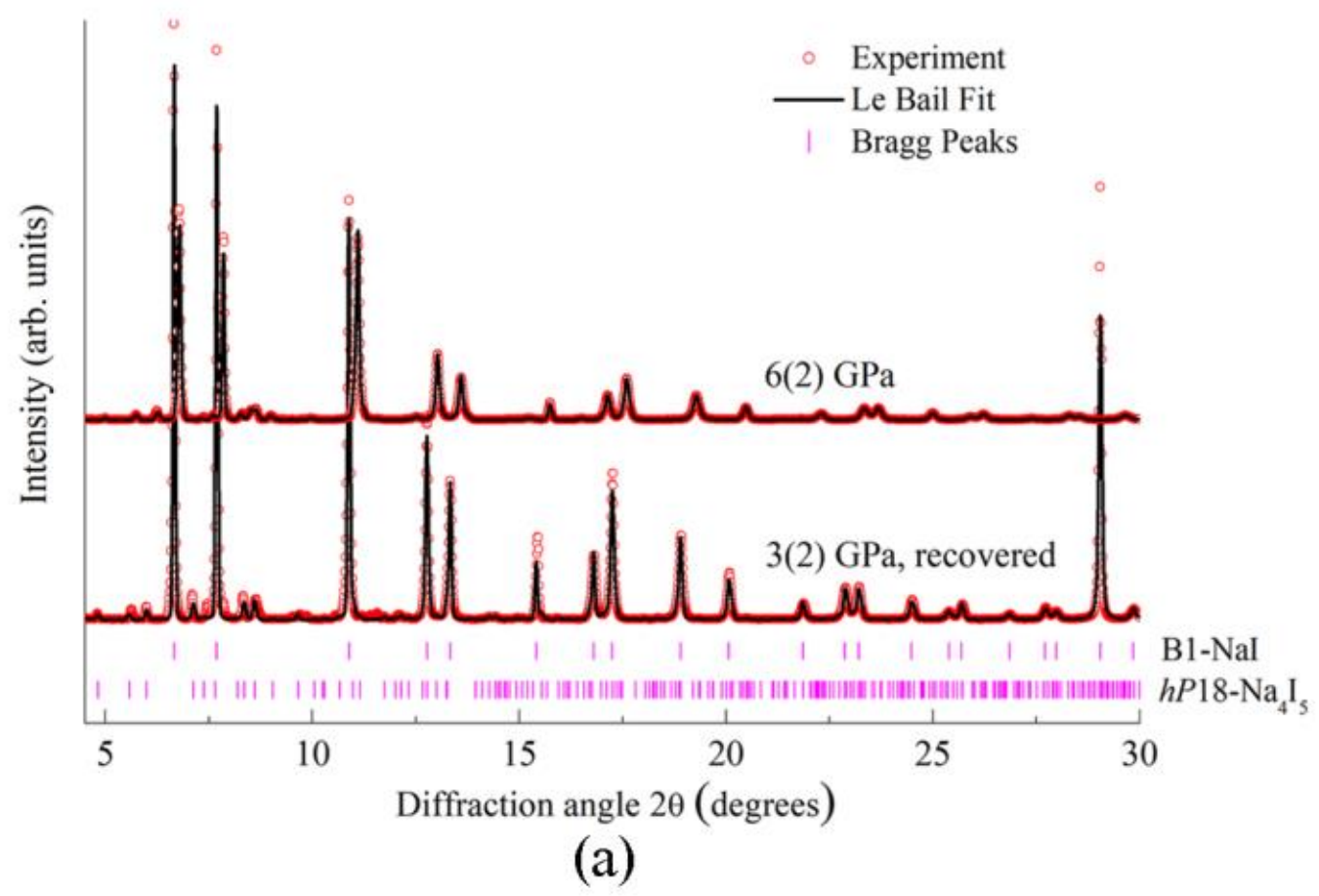

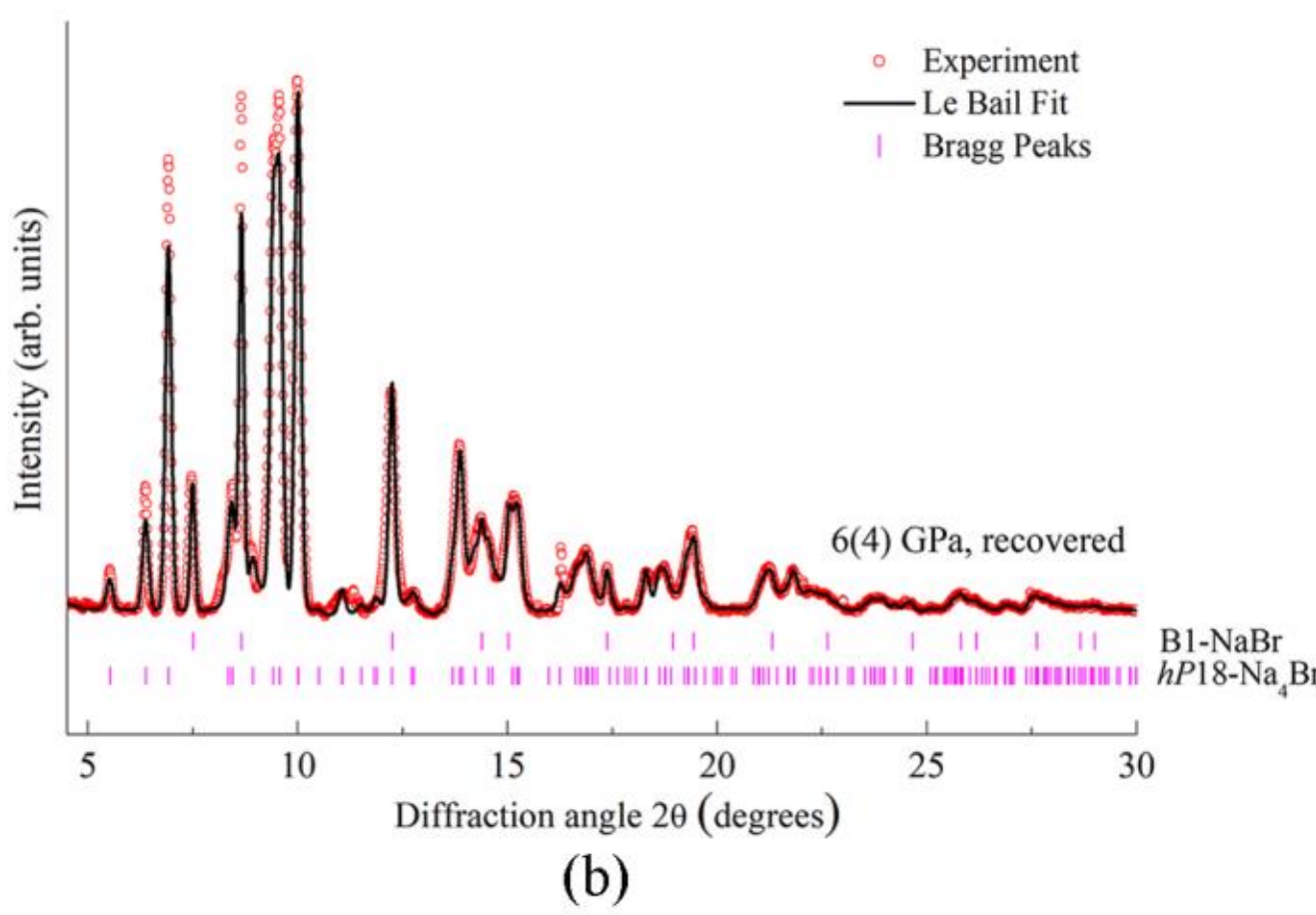

**Fig. S2.** Powder XRD patterns obtained from various samples in three different DACs after the synthesis of (a) *hP*18-$Na_4I_5$ and (b) *hP*18-$Na_4Br_5$. (a) the sample in DAC #1 upon decompression at 6(2) GPa; the sample in DAC #2 at 3(2) GPa (notated “recovered”, as it was first recovered to ambient pressure in the DAC opened in a glove bag and then closed again; (b) the sample in DAC #5 at 6(4) GPa (also after recovery). Vertical ticks correspond to Bragg peaks of B1-NaBr, *hP*18-$Na_4Br_5$, B1-NaI and *hP*18-$Na_4I_5$; red open circles are experimental points; black lines- the Le Bail fit. X-ray wavelength $\lambda = 0.4099$ Å.

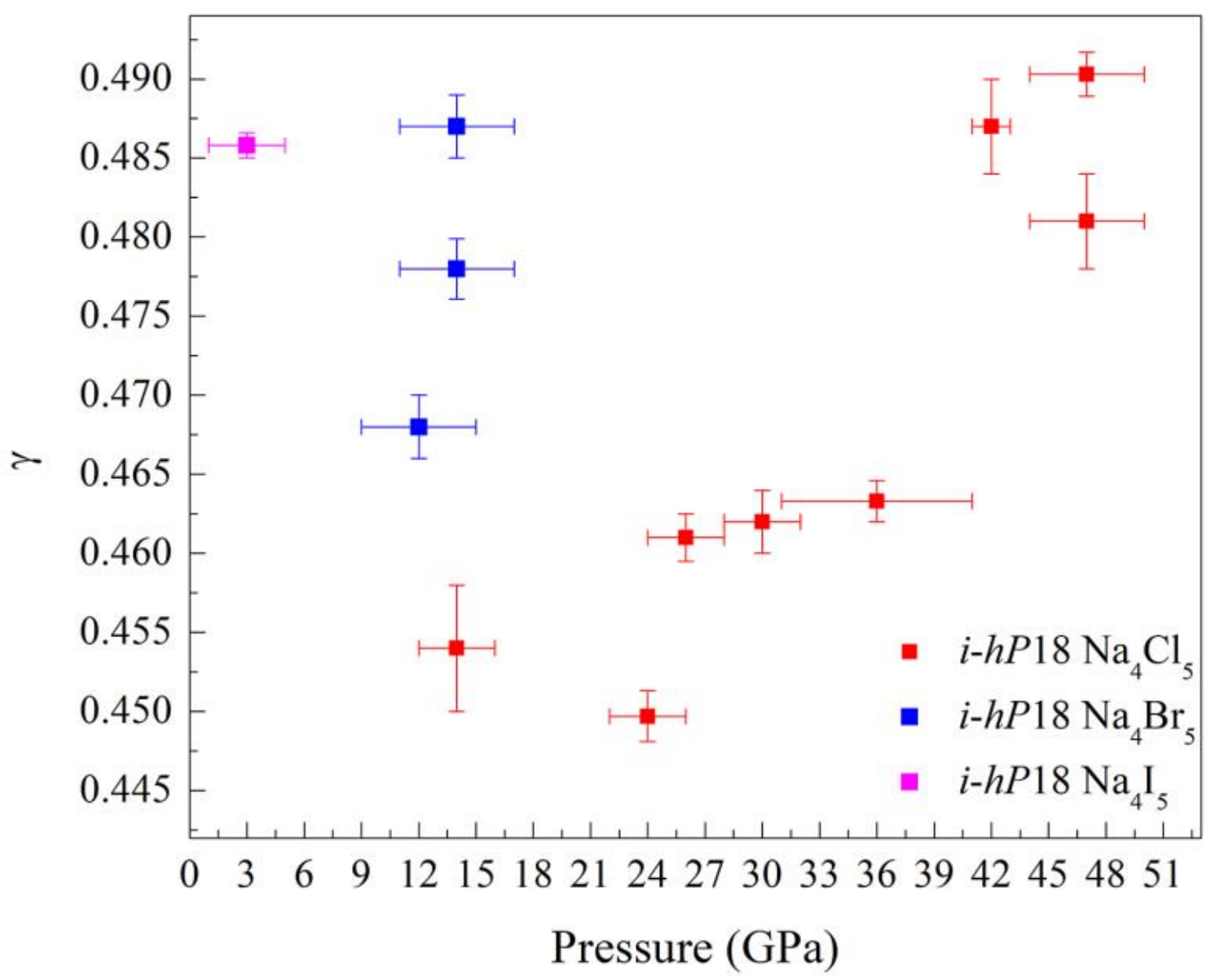

**Fig. S3.** Pressure dependence of the γ component of the modulation wavevector **q** = γ**c*** in the incommensurate *i-hP*18 $Na_4X_5$ phases.

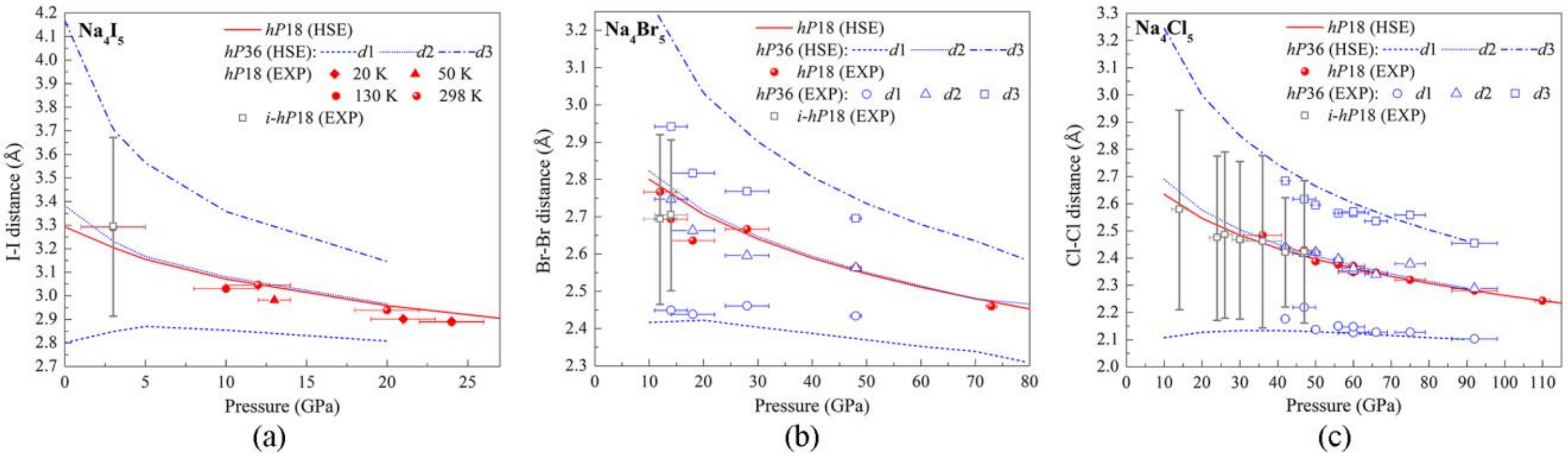

**Fig. S4.** Comparison of experimental X-X distances in (a) $Na_4I_5$, (b) $Na_4Br_5$, and (c) $Na_4Cl_5$ with those calculated with the HSE06 functional. The grey squares represent the refined average value of the X-X distances in chains of the *i-hP*18, and the X-X distances are continuously distributed between the error bars.

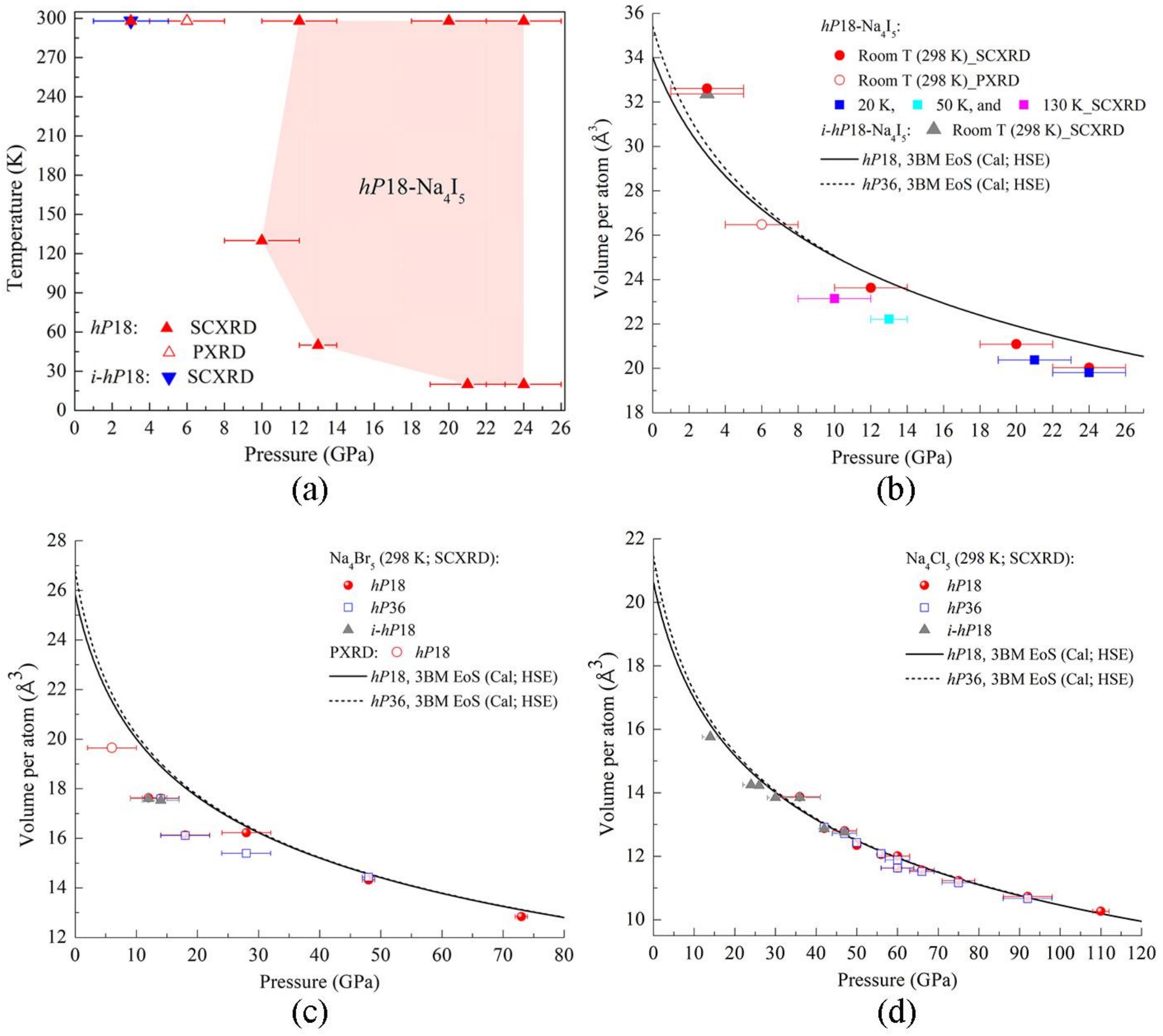

**Fig. S5.** (a) Experimental pressure-temperature (P-T) diagram of the $Na_4I_5$ compound. (b), (c), and (d) The volume per atom as a function of pressure from our DFT calculations using the HSE06 functional in comparison with experimental data for $Na_4I_5$, $Na_4Br_5$, and $Na_4Cl_5$, respectively. Lines represent DFT-calculated volume for the given pressure fitted by the third-order Birch-Murnaghan equation of state (see parameters in Table S9).

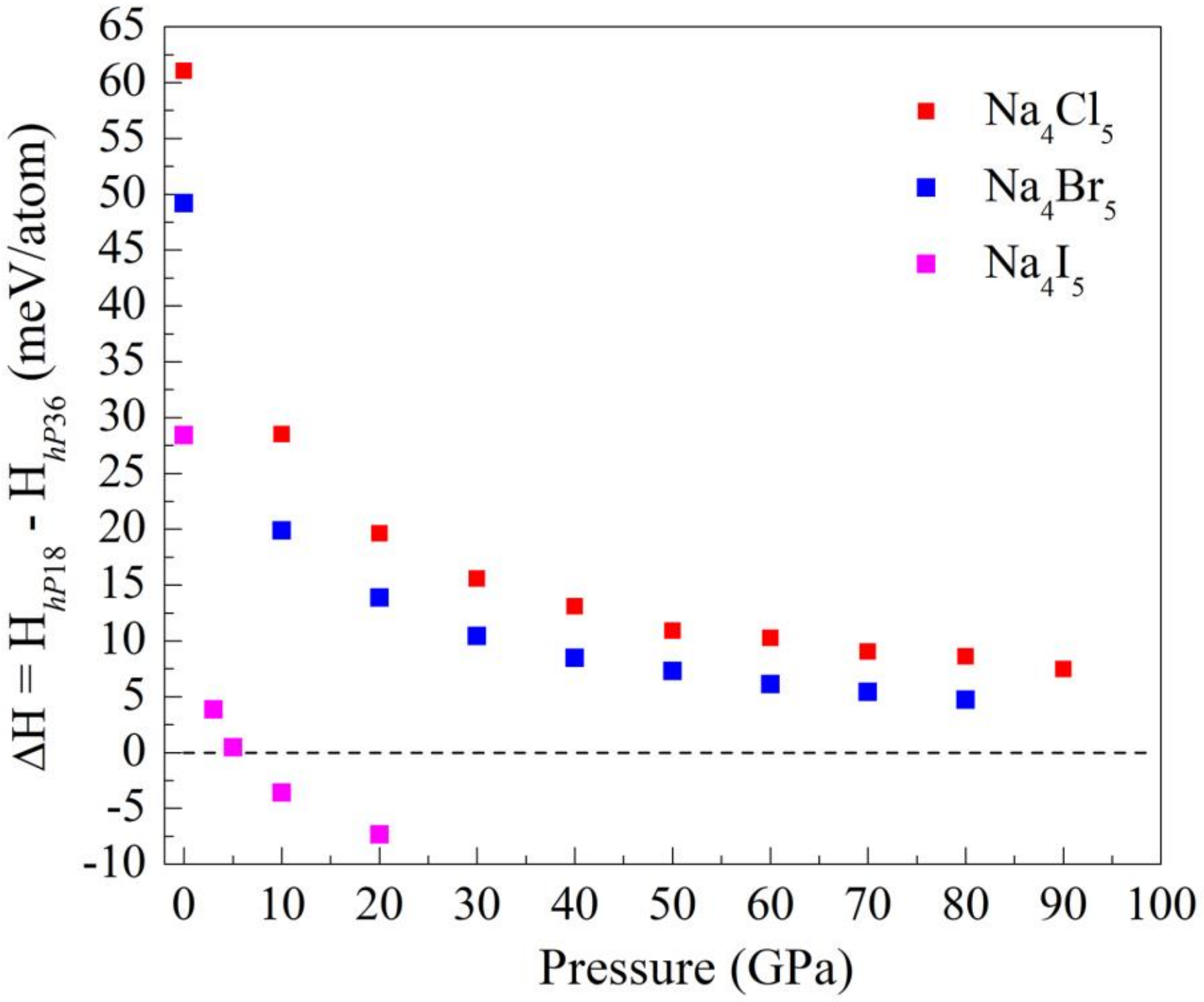

**Fig. S6.** Enthalpy difference between the *hP*18- and *hP*36-$Na_4X_5$ ($\Delta H = H_{hP18} - H_{hP36}$) calculated at zero temperature using the HSE06 functional. The enthalpy differences may not show phase transitions because of dynamical instability of *hP*18-compounds at T = 0 K, but they reflect the stability trends as a function of pressure and atomic number of X atoms. All values are smaller than the thermal energy contribution of ~192 meV/atom ($k_BT$ at ~2200 K).

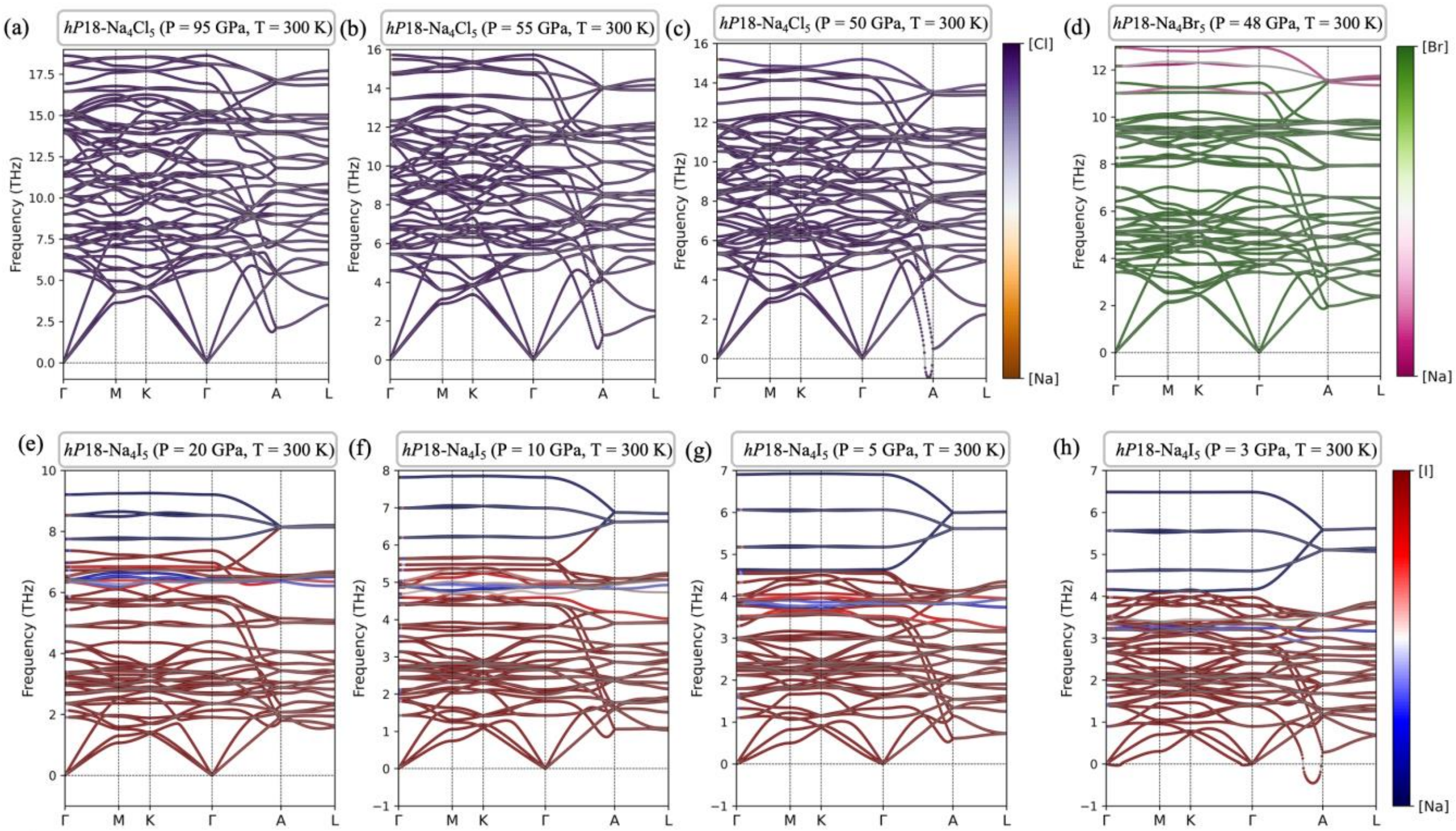

**Fig. S7.** Phonon dispersion relations of *hP*18-$Na_4Cl_5$, *hP*18-$Na_4Br_5$, and *hP*18-$Na_4I_5$ at room temperature ($T$ = 300 K) for the pressure points corresponding to the triangular symbols in the diagram shown in Fig. S8. Panels (a–c) show the phonon dispersion relations of *hP*18-$Na_4Cl_5$ at 50, 55, and 95 GPa, respectively. Panel (d) presents the phonon dispersion relations of *hP*18-$Na_4Br_5$ at 48 GPa. Panels (e–h) represents the phonon dispersion relations for *hP*18 $Na_4I_5$ calculated at pressures ranging from 20 GPa to 3 GPa, corresponding to the experimental synthesis conditions in the phase diagram. For dynamically stable phases, all vibrational frequencies are real, while the presence of imaginary frequencies (shown in the figure as negative frequencies) indicates dynamical instability. The soft modes observed near A point along the Γ–A direction in *hP*18-$Na_4X_5$ indicate a transition to incommensurately modulated structures. The calculated wave vectors for the soft modes are q = 0.48 $Å^{-1}$ for *hP*18-$Na_4Cl_5$ at 50 GPa, q = 0.479 $Å^{-1}$ for *hP*18-$Na_4Br_5$ at 48 GPa and q = 0.467 $Å^{-1}$ for *hP*18-$Na_4I_5$ at 3 GPa. For the experimentally observed incommensurately modulated structures with the space group $P6_3/mcm(00\gamma)00ss$ $\gamma$ = 0.481(3) for *i-hP*18-$Na_4Cl_5$ at 47(3) GPa, $\gamma$ = 0.4780(19) for *i-hP*18-$Na_4Br_5$ at 14(3) GPa and $\gamma$ = 0.4858(8) for *i-hP*18-$Na_4I_5$ at 3(2) GPa. This result shows excellent agreement with our experimentally observed incommensurately modulated *i-hP*18 phase of $Na_4X_5$ (see also Fig. S8).

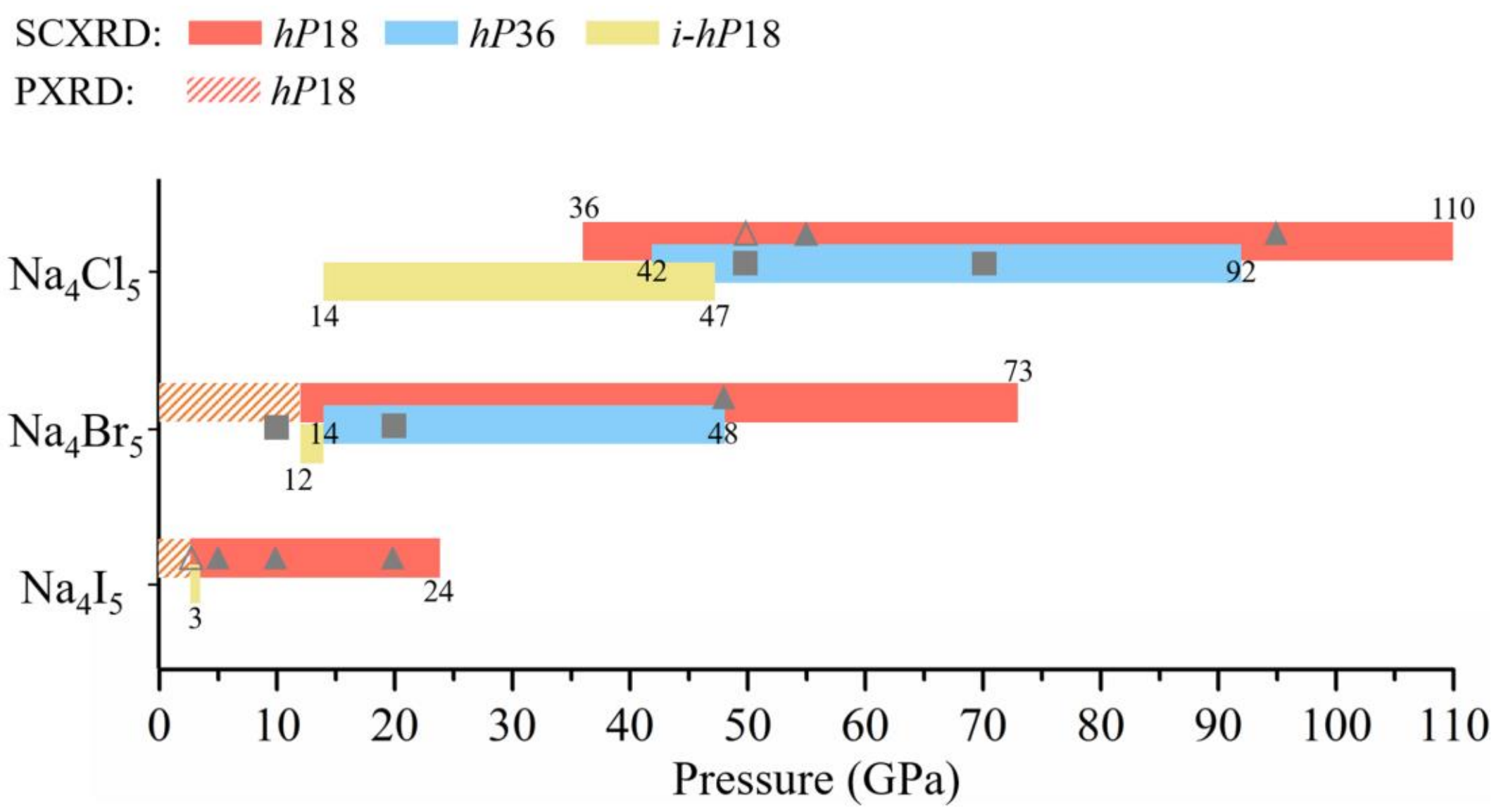

**Fig. S8.** Schematic diagram presenting pressure intervals in which $Na_4X_5$ phases were experimentally observed at room temperature, along with the results of theoretical evaluation of the dynamic stability of *hP*18 and *hP*36 compounds. The stability evaluation is based on our calculated phonon dispersion relations: at finite temperature (300 K) for the *hP*18 phases and at 0 K for the *hP*36 phases at different pressures (see Fig. S7 and Fig. S9). Solid and opaque triangles (squares) denote dynamically stable and dynamically unstable *hP*18 (*hP*36) compounds, respectively.

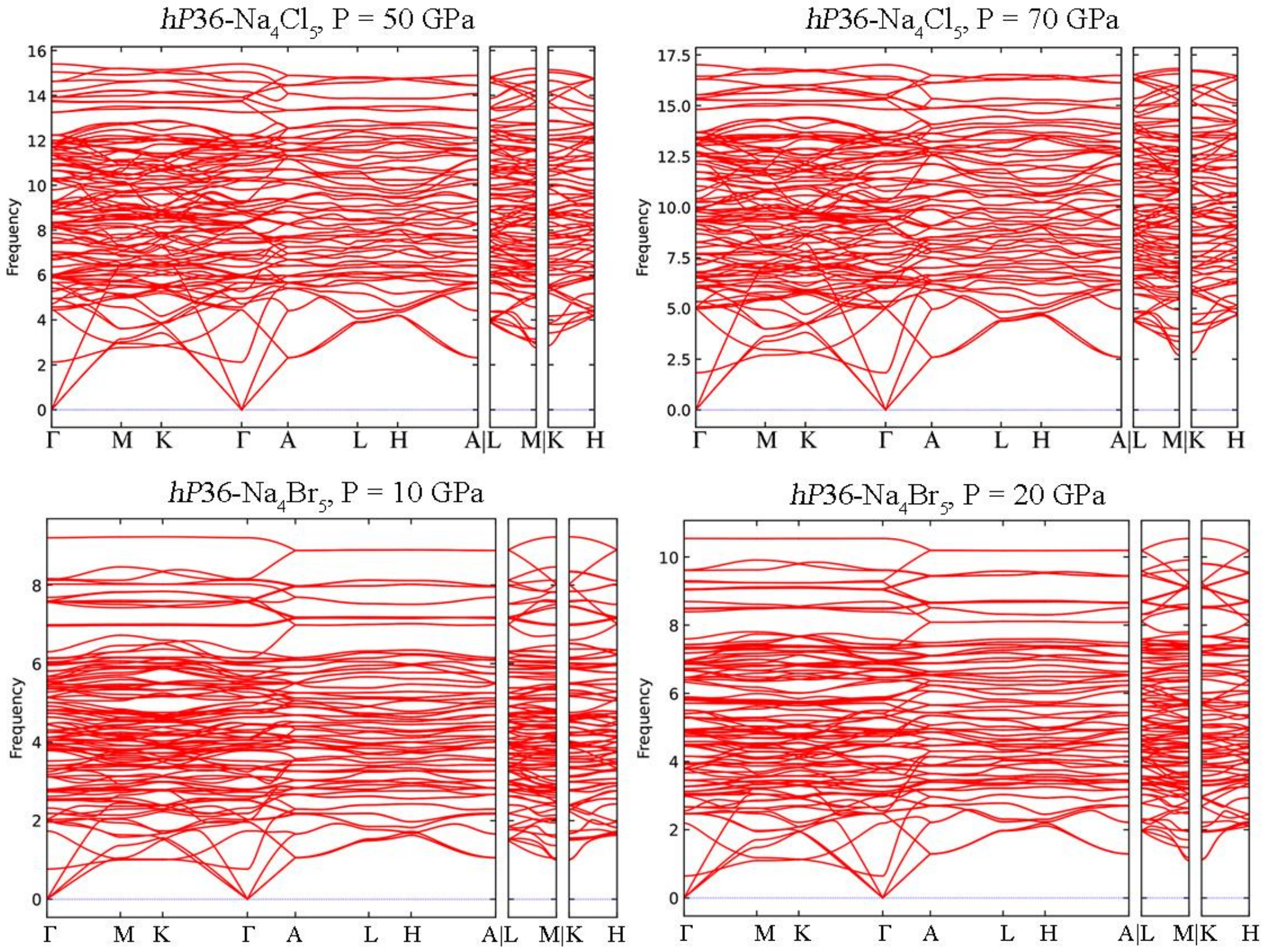

**Fig. S9.** Phonon dispersion curves of *hP*36-$Na_4Cl_5$ and *hP*36-$Na_4Br_5$ calculated using HSE06 along high-symmetry directions in the Brillouin zone at selected pressures.

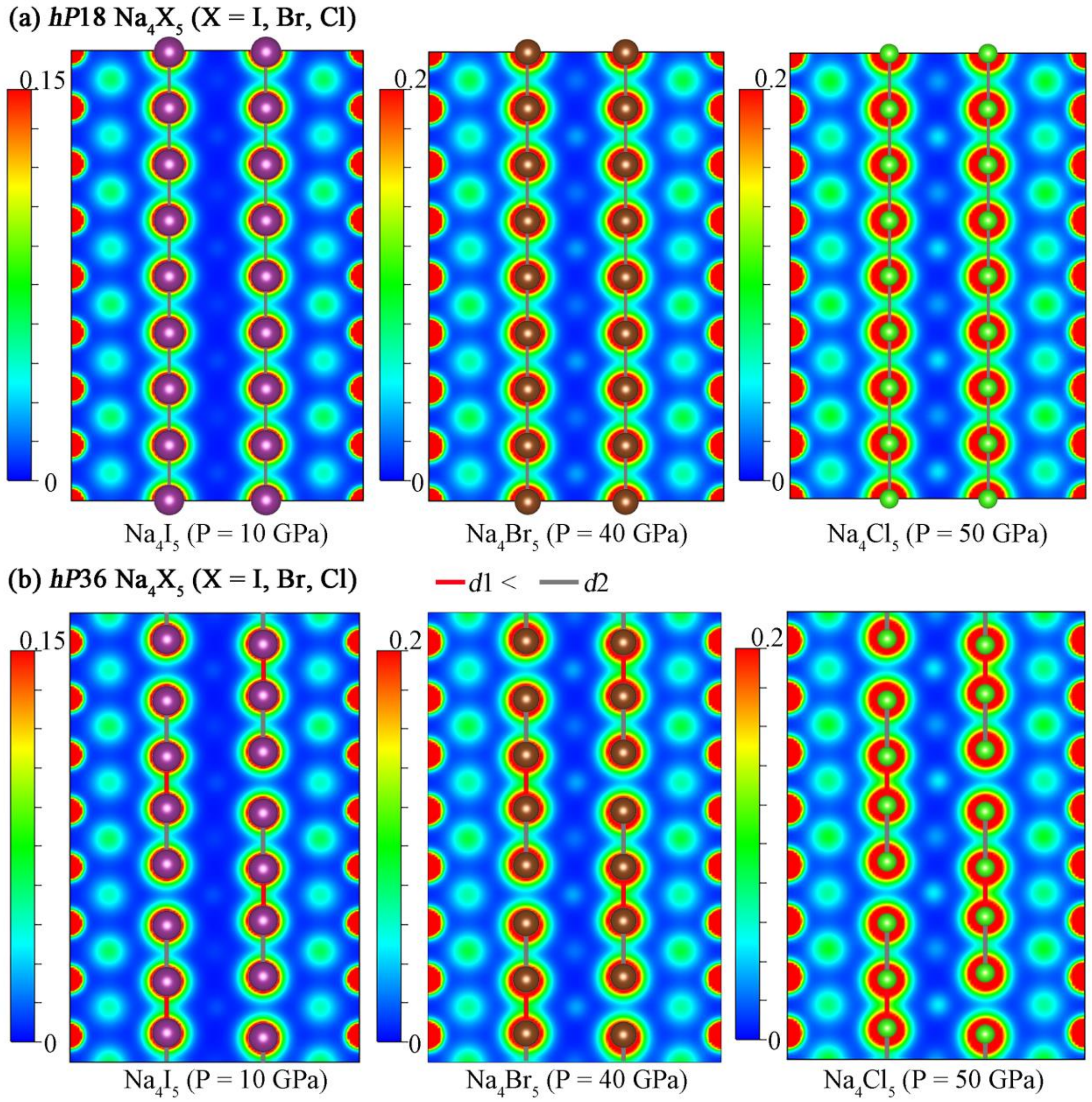

**Fig. S10.** Calculated charge density maps in the (1 1 0) plane of (a) *hP*18 $Na_4X_5$ and (b) *hP*36 $Na_4X_5$ using the hybrid (HSE06) functional at different pressures. The charge density maps clearly depicts that the 1D halide chains are electronically isolated from the surrounding bulk subsystems. A strong electronic overlap is observed along the chains oriented along the *c*-axis of *hP*18 phases, whereas in *hP*36 $Na_4X_5$ the formation of linear tetrads (linear $X_4^{2-}$ units) is observed. The color scale is in e/bohr$^3$.

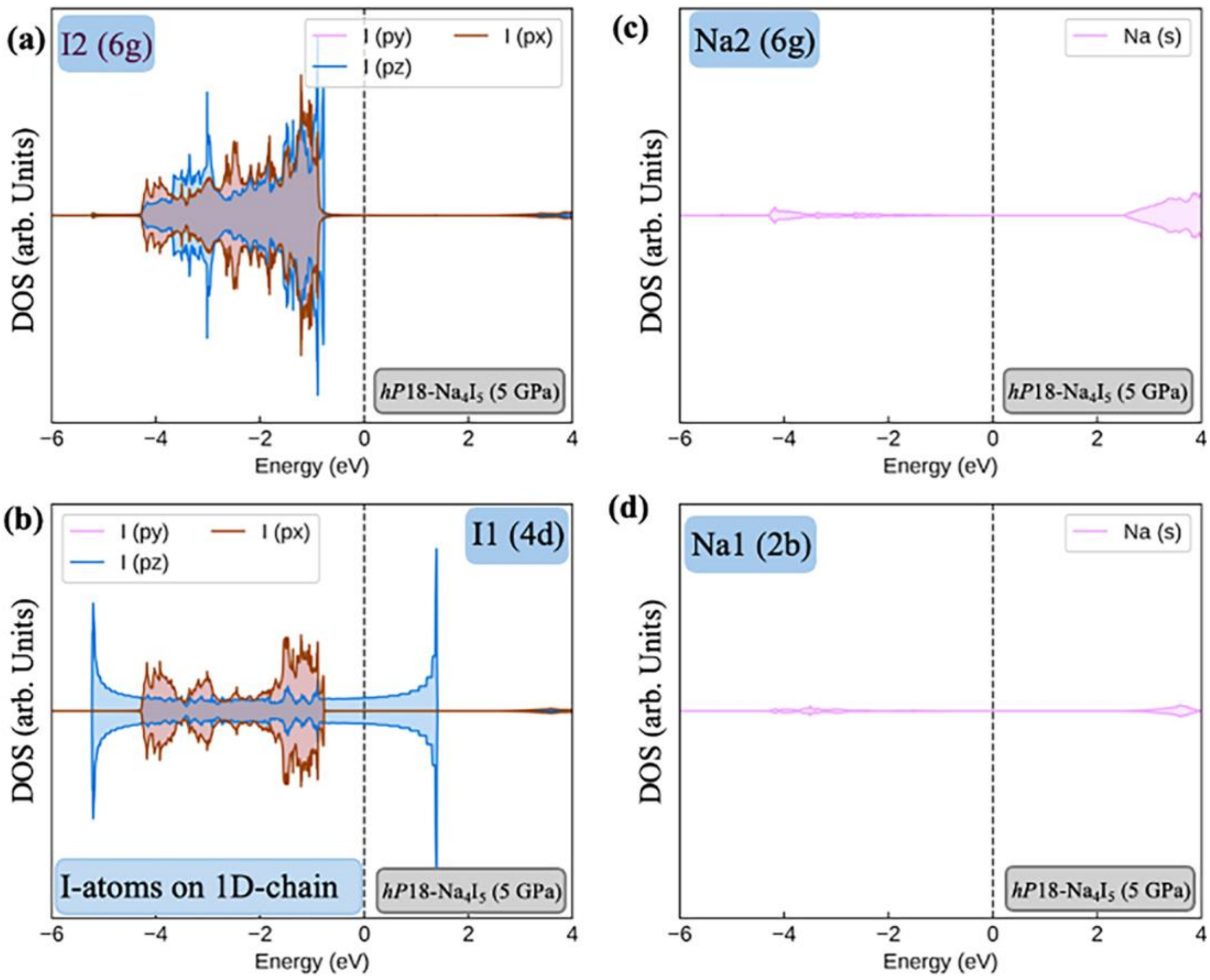

**Fig. S11.** Atom and orbital projected electronic DOS of *hP*18-$Na_4I_5$ at 5 GPa. The projections have been made onto different Wyckoff sites of cations (Na1, Na2) and anions (I1, I2) in order to show that the electronic density around the Fermi energy is mainly arising from the I1 atoms in 4*d* Wyckoff site due to a strong overlap of I ($5p_z$) orbitals in the linear I-I chains.

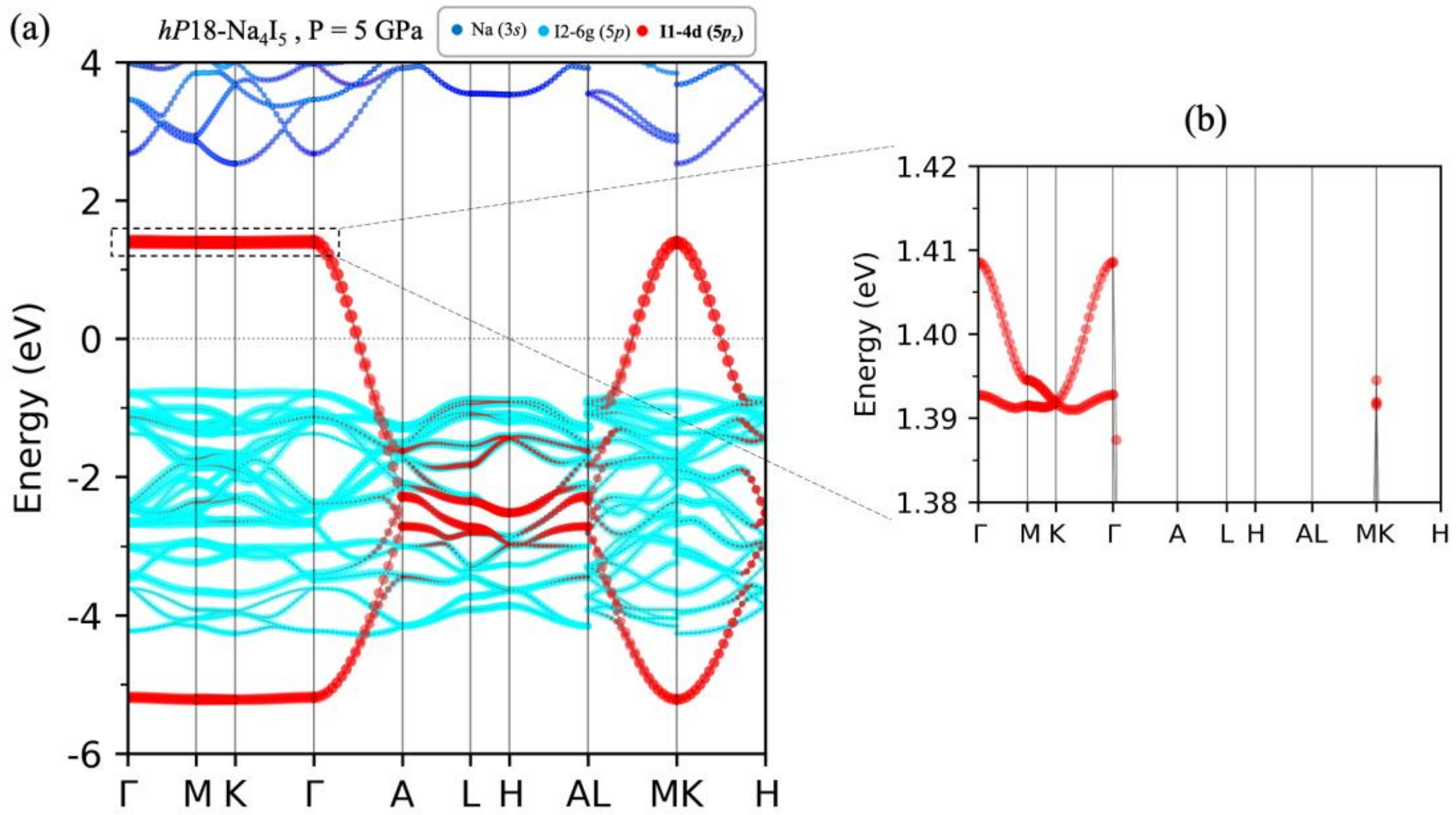

**Fig. S12.** (a) Atom and orbital projected electronic band structure of *hP*18-$Na_4I_5$ at 5 GPa in a static lattice (T = 0 K). The projected electronic bands (in red color) arising from the I1 atoms at 4*d* Wyckoff site and consist of overlapping I-5$p_z$ orbitals along the linear chain. (b) A zoomed section of the flat-bands along the Γ–M–K–Γ path in the reciprocal space depicting a linear band crossing at the K-point.

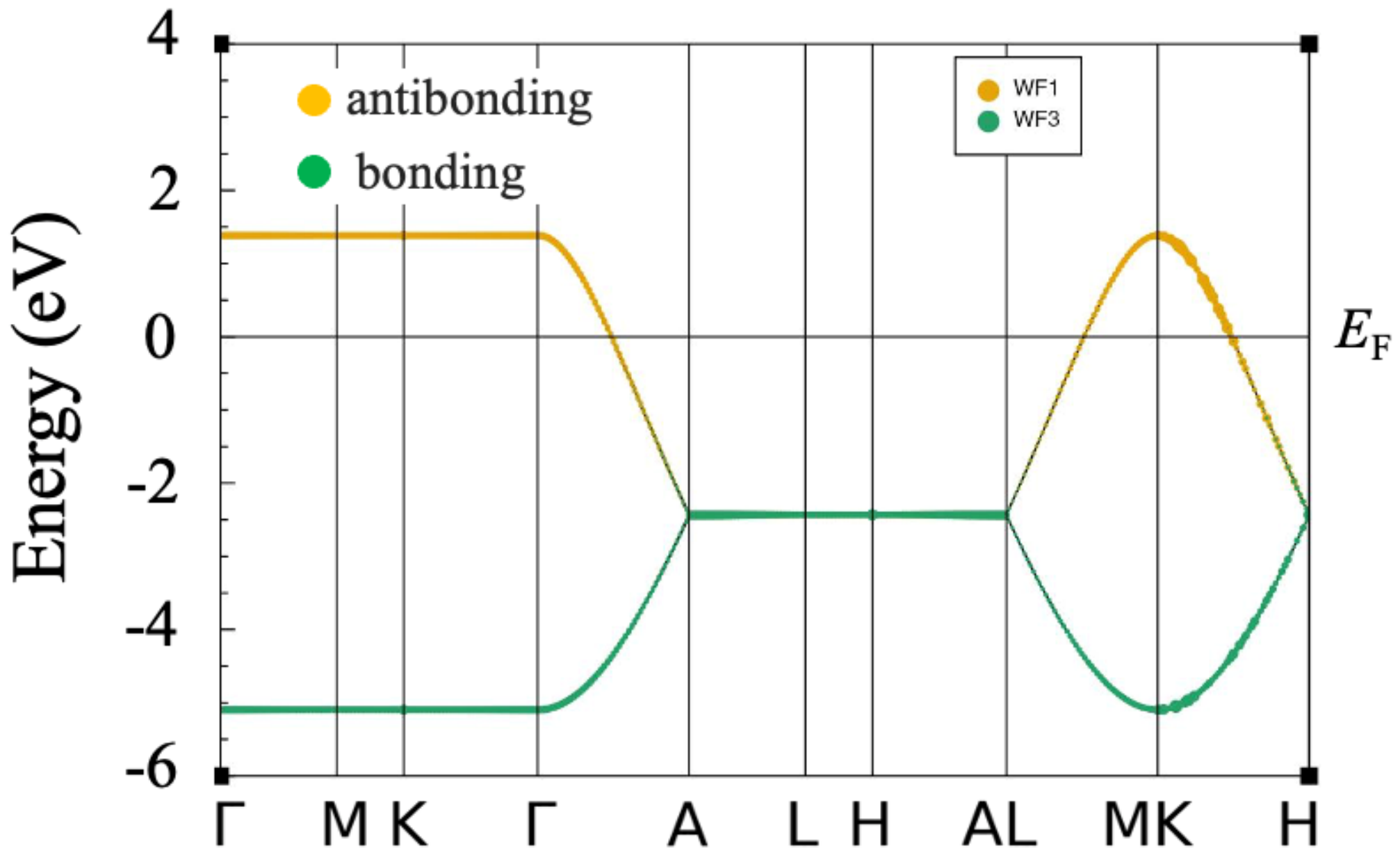

**Fig. S13.** Wannier interpolated band structure considering the isolated 1D chains of iodine atoms in the unit cell of *hP*18-$Na_4I_5$.

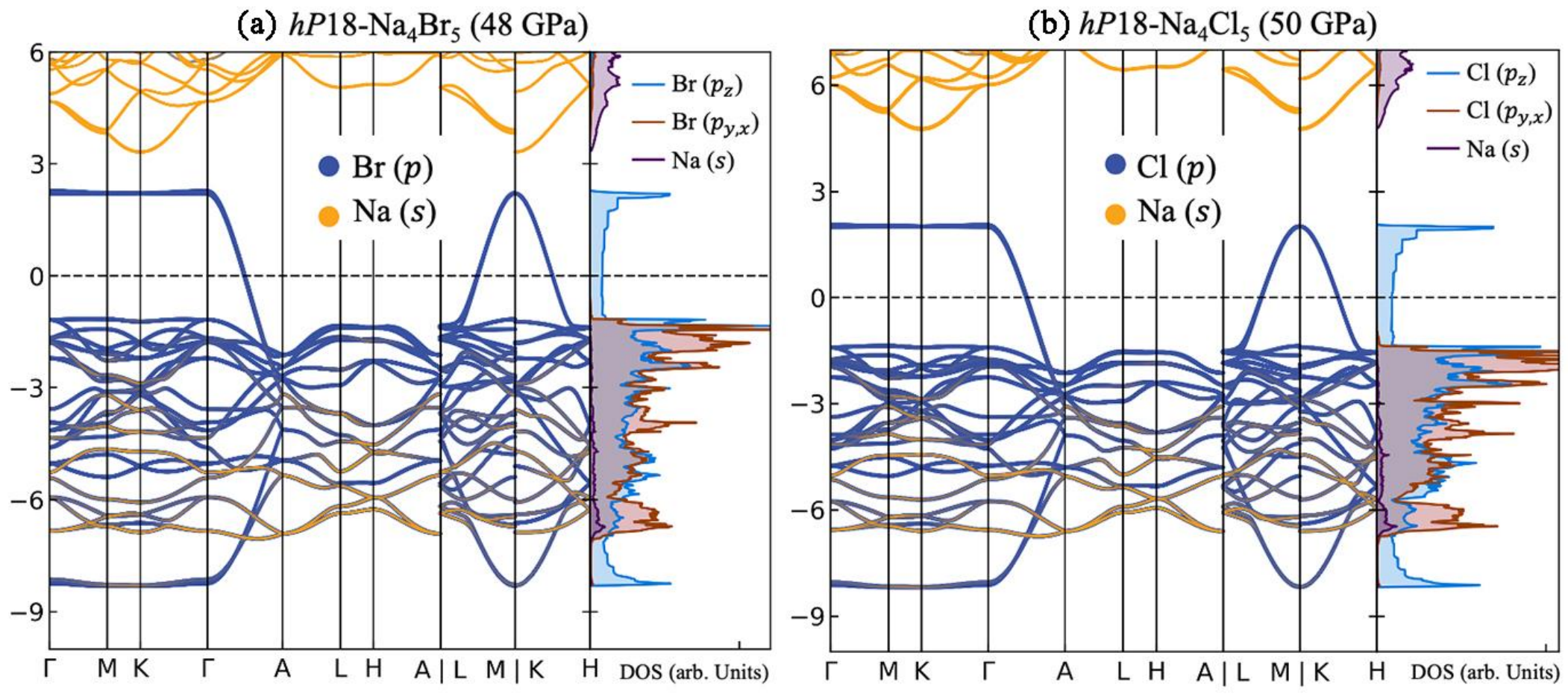

**Fig. S14.** Electronic band structure and the corresponding density of states (DOS) for (a) *hP*18 $Na_4Br_5$ and (b) $Na_4Cl_5$ at pressures of 48 GPa and 50 GPa, respectively, calculated with GGA-PBE functional considering a static (T = 0 K) lattice arrangement.

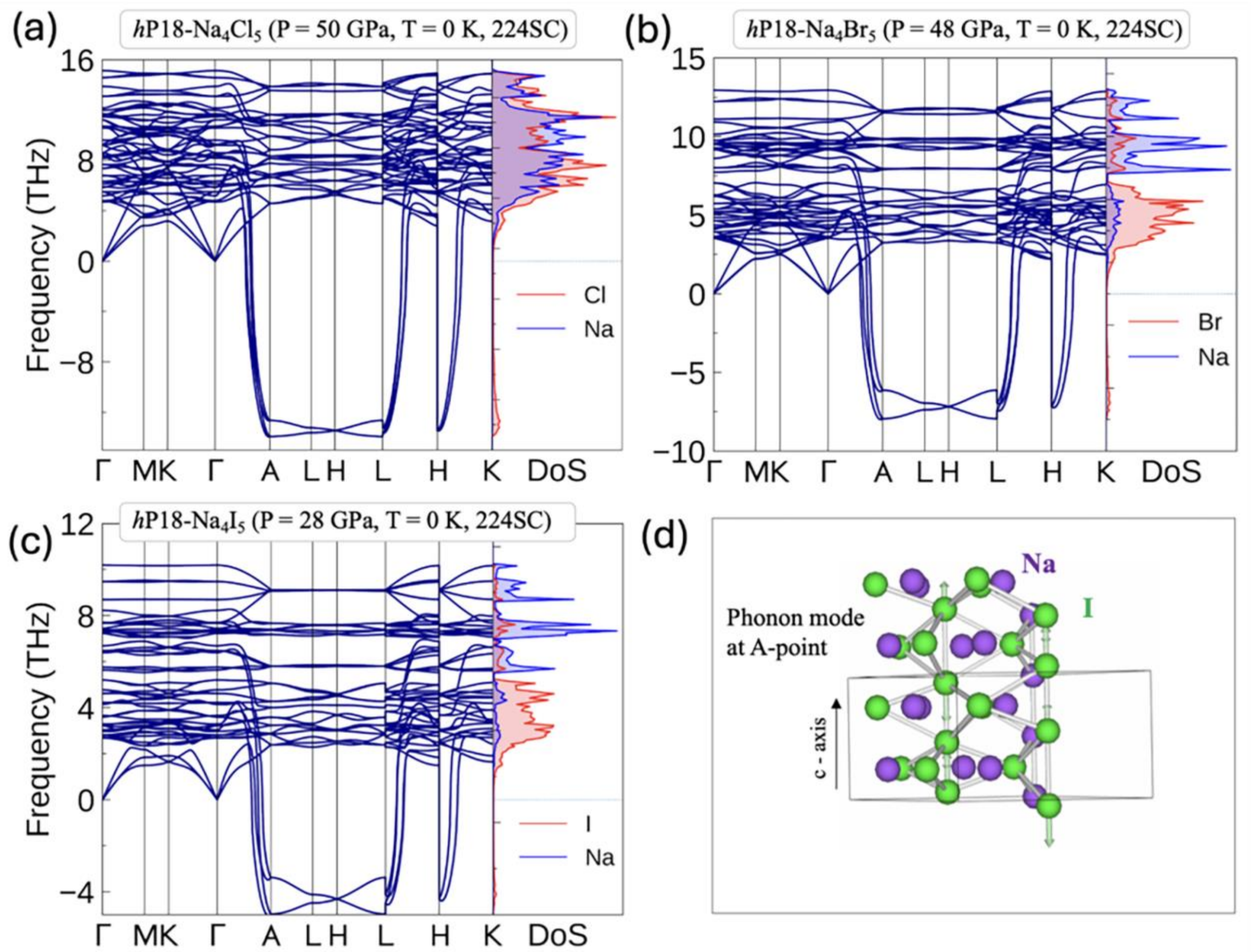

**Fig. S15.** Phonon dispersion relations for the *hP*18 $Na_4Cl_5$, $Na_4Br_5$, and $Na_4I_5$ calculated selected pressures of 50, 48, and 28 GPa, respectively, within the harmonic approximation (T = 0 K). When considered an incommensurate supercell (SC) size of 2×2×3, all phonon modes were found to be real using harmonic approximation (*21*), indicating lattice dynamic stability. Using a commensurate 2×2×4 SC reveal a pronounced imaginary phonon branch along the Γ-A path, eigenvector corresponds to the X-X chain direction (i.e., along the *c* axis), in agreement with what is expected for Peierls materials (*11*). Visualization of the phonon eigenvector at the A-point of the wavevector shows that the unstable mode corresponds to correlated vibrations of halogen atoms along the 1D chain.

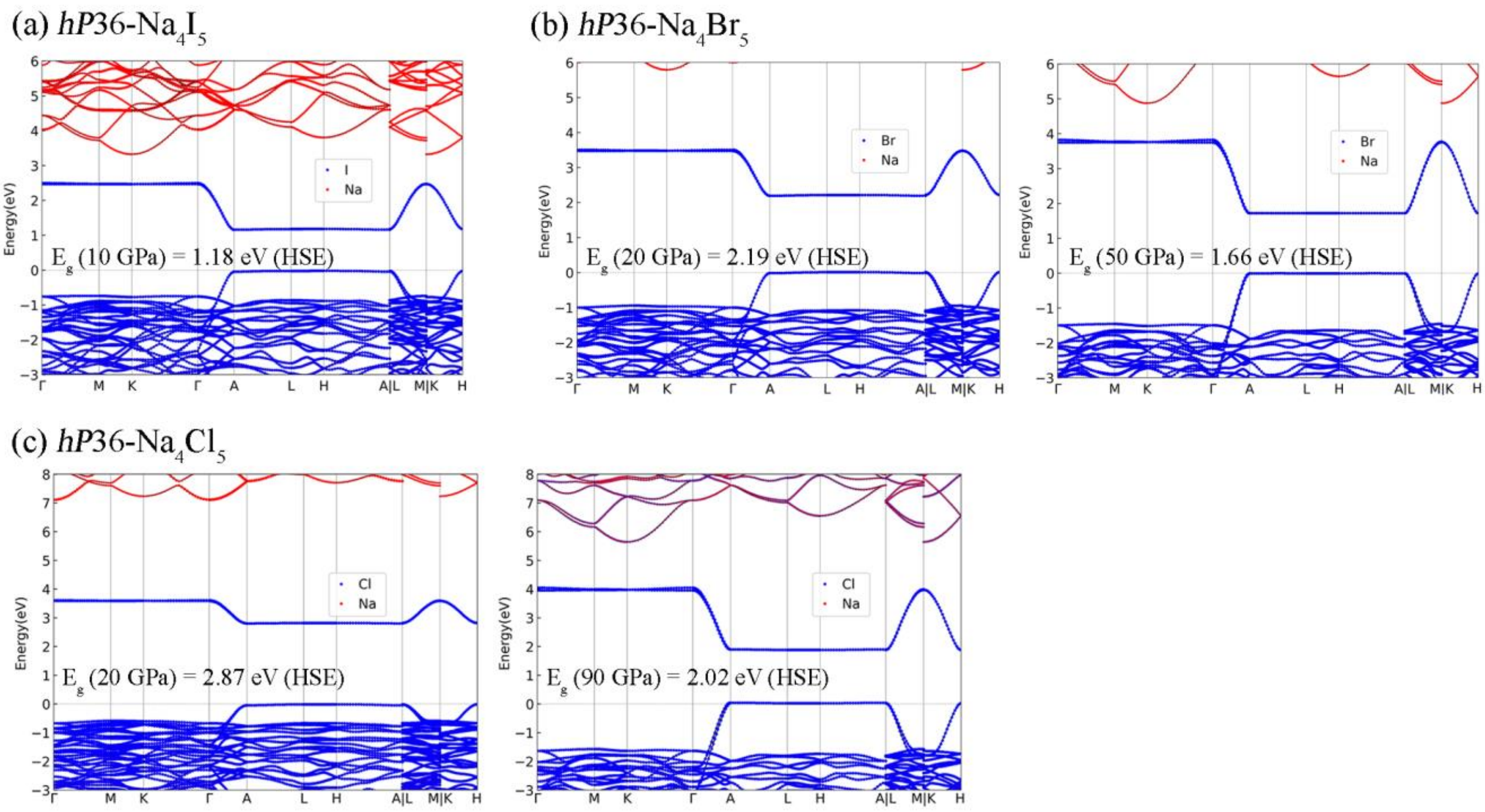

**Fig. S16.** Calculated band structures using hybrid (HSE06) functional for (a) *hP*36-$Na_4I_5$, (b) *hP*36-$Na_4Br_5$, and (c) *hP*36-$Na_4Cl_5$ at different pressures.

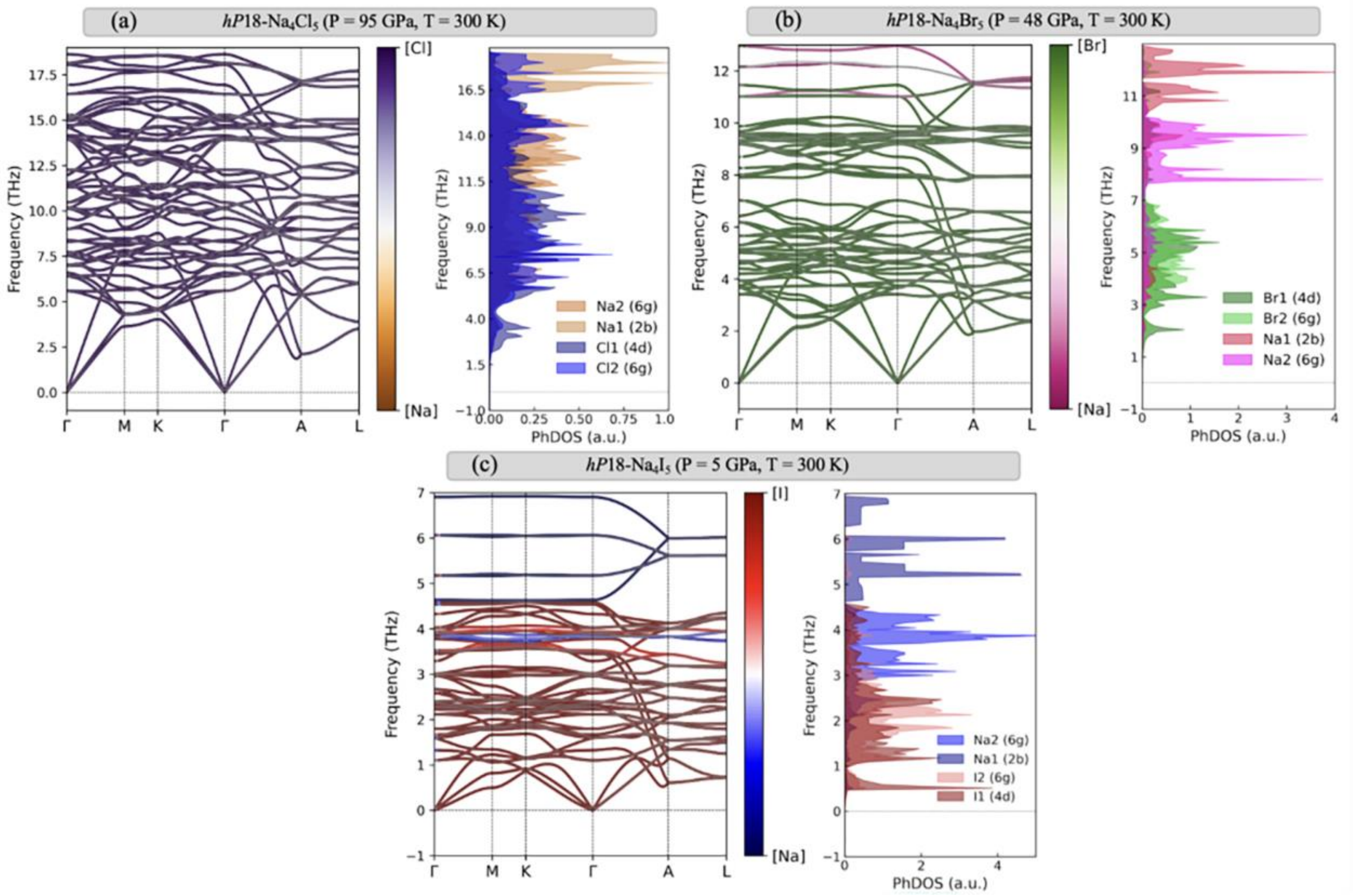

**Fig. S17.** Phonon dispersion relations at 300 K, and projected phonon density of states (PhDOS) for *hP*18 $Na_4X_5$ (X= Cl, Br, I) calculated at selected pressures. The X1 atoms at the 4*d* Wyckoff site corresponds to the halide atoms vibrating on the linear 1D chains.

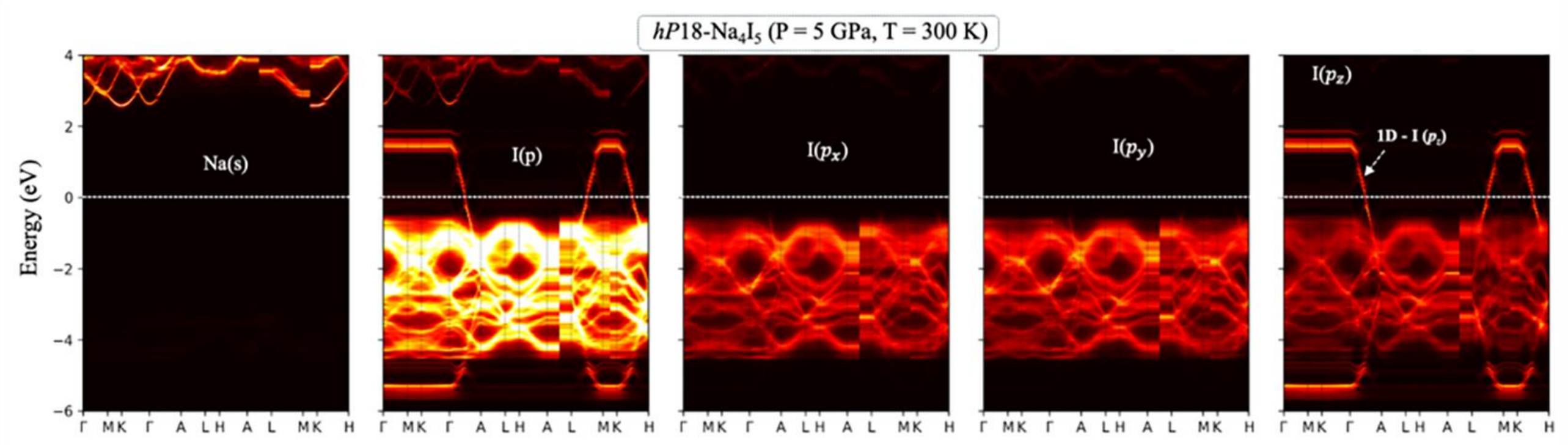

**Fig. S18.** Atom and orbital resolved effective band structures (EBS) of *hP*18-$Na_4I_5$ at 5 GPa and 300 K. The orbital resolved EBS depict that the 1D electronic bands are mainly due to the I (5$p_z$) orbitals from the linear I chains.

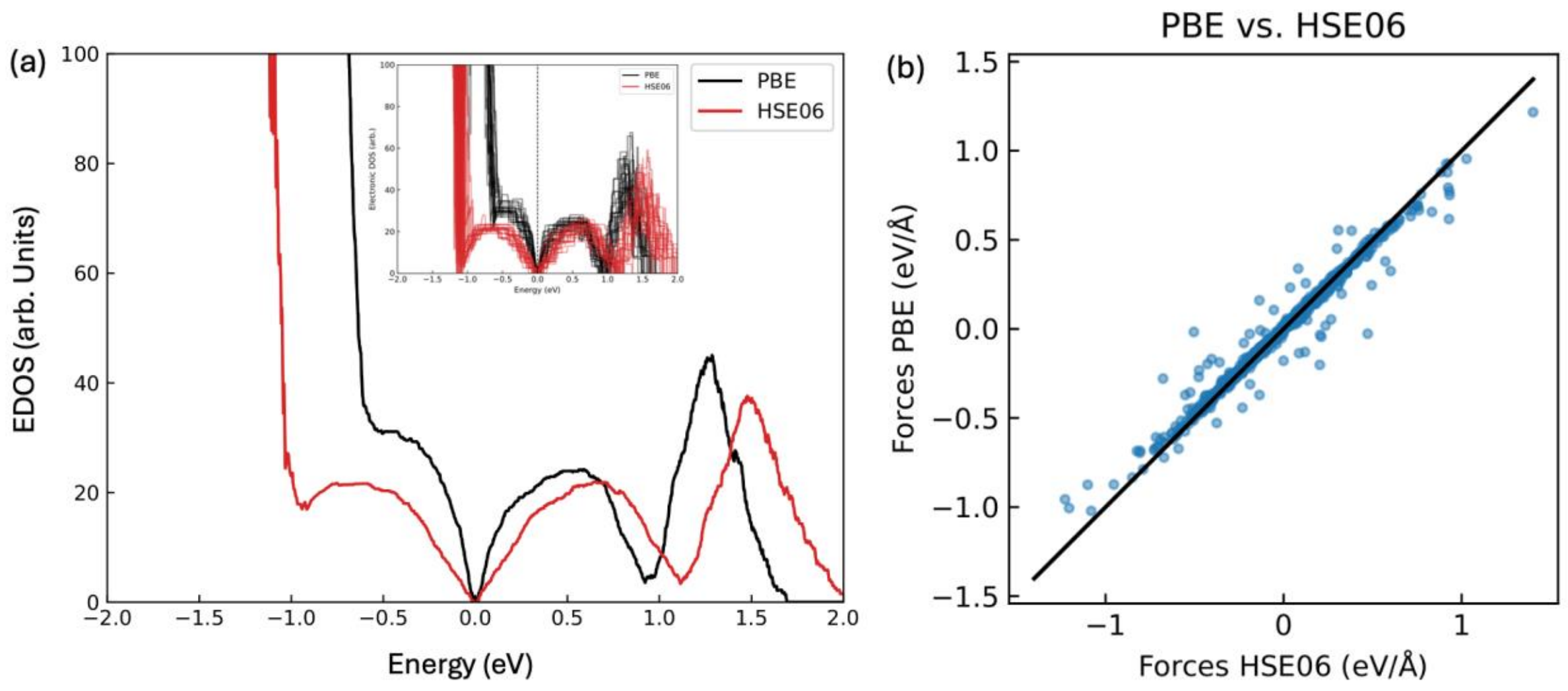

**Fig. S19.** (a-b) Comparison of the averaged electronic density of states (left) and HF forces (right) calculated for a set of 20 uncorrelated AIMD samples (at 300 K) with PBE GGA and HSE DFT for *hP*18-$Na_4I_5$ at 5 GPa. The figure inset in (a) includes the unaveraged DOS from 20 thermalized samples.

## Supplementary Discussion 1

Reconstructed reciprocal lattice planes of $Na_4X_5$ at selected pressures

1. $Na_4I_5$. SCXRD collected at pressures above 3(2) GPa over a broad temperature range (Table S1) consistently refines to the *hP*18 lattice and shows no superlattice reflections (Fig. SD1. 1a, b). At 3(2) GPa, some domains remain indexable with the same *hP*18 cell, yet some domains exhibit weak additional reflections at the middle of *c** (Fig. SD1. 1c). One-dimensional line-profile analysis across these spots reveals clear peak splitting (Fig. SD1. 1d). These satellite reflections can be indexed with a modulation wavevector $\mathbf{q} = \gamma\mathbf{c^*}$, where $\gamma = 0.4858$.

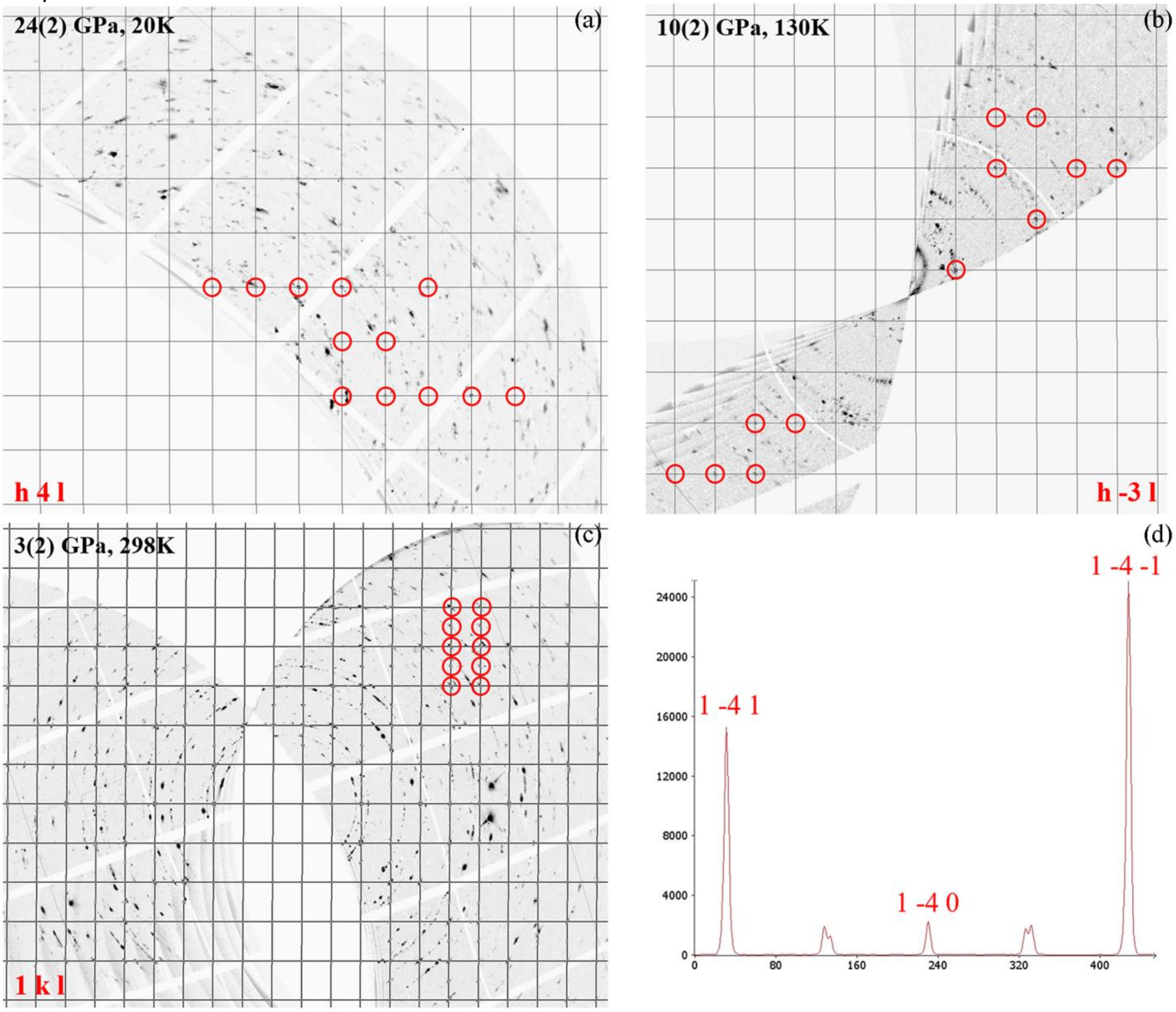

Fig. SD1. 1. Reconstructed reciprocal lattice planes of $Na_4I_5$ at (a) 24(2) GPa, 20 K, (b) 10(2) GPa, 130 K, and (c) 3(2) GPa, 298 K from the experimental SCXRD dataset using CrysAlis$^{Pro}$ software. A one-dimensional line-profile across some selected reflections is shown in (d), highlighting the satellite peak splitting. The data was collected at ID15b beamline at ESRF with a beam size of 2×2 μm$^2$.

2. $Na_4Br_5$. Similarly, the below Fig. SD1. 2 reveals coexist of the three phases of $Na_4Br_5$ at 14(3) GPa within one collection spot (beam size of 2×2 μm$^2$).

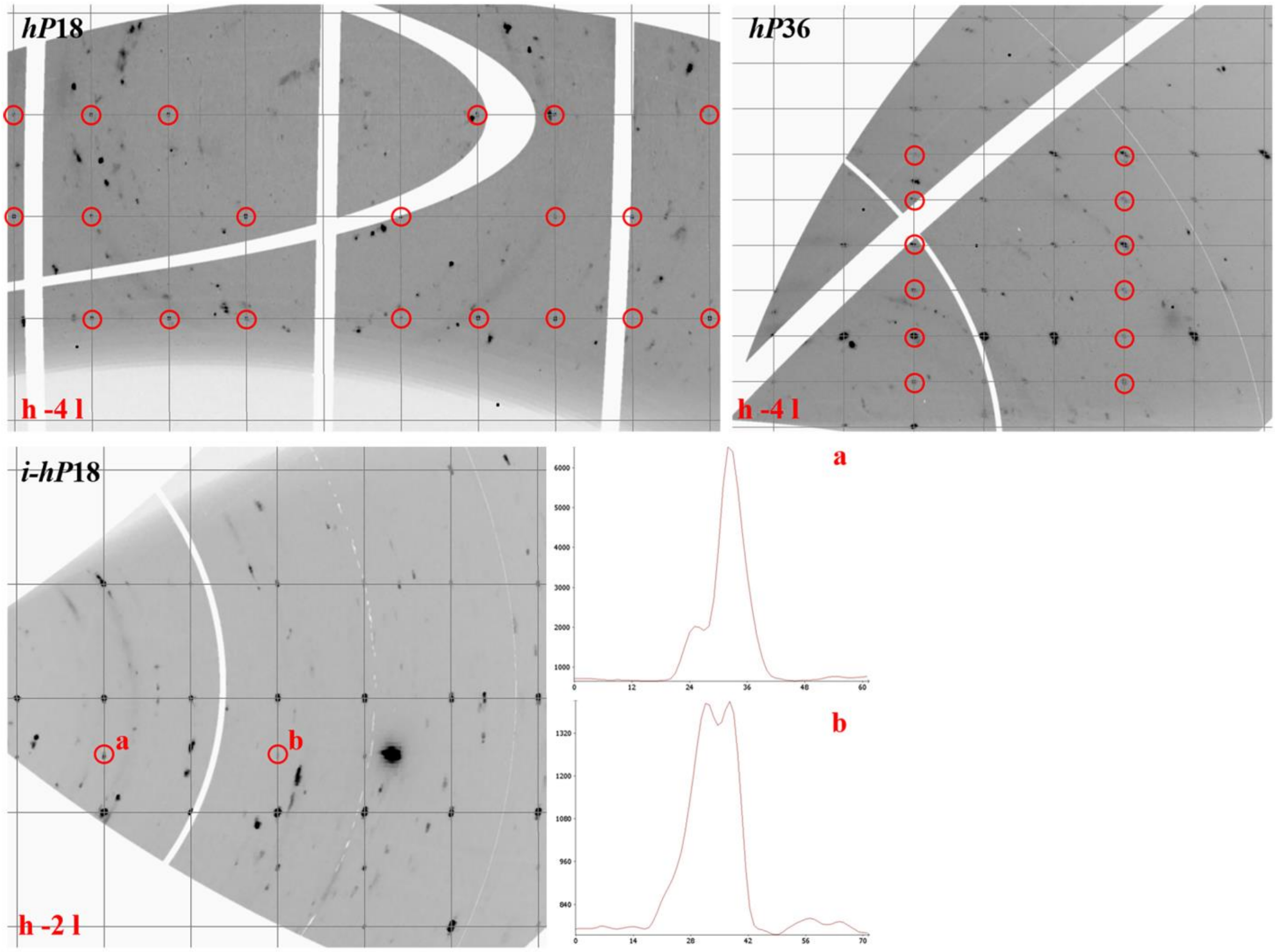

Fig. SD1. 2. Reconstructed reciprocal lattice planes of $Na_4Br_5$ at 14(3) GPa on different domains of *hP*18, *hP*36, and *i-hP*18 from the experimental SCXRD dataset using CrysAlis$^{Pro}$ software. One-dimensional line-profiles across the selected reflections a and b of *i-hP*18 are shown, highlighting the satellite peak splitting. The data was collected at ID15b beamline at ESRF with a beam size of 2×2 μm$^2$.

3. $Na_4Cl_5$. Similarly, the below Fig. SD1. 3 reveals the three phases of $Na_4Cl_5$ at different pressures.

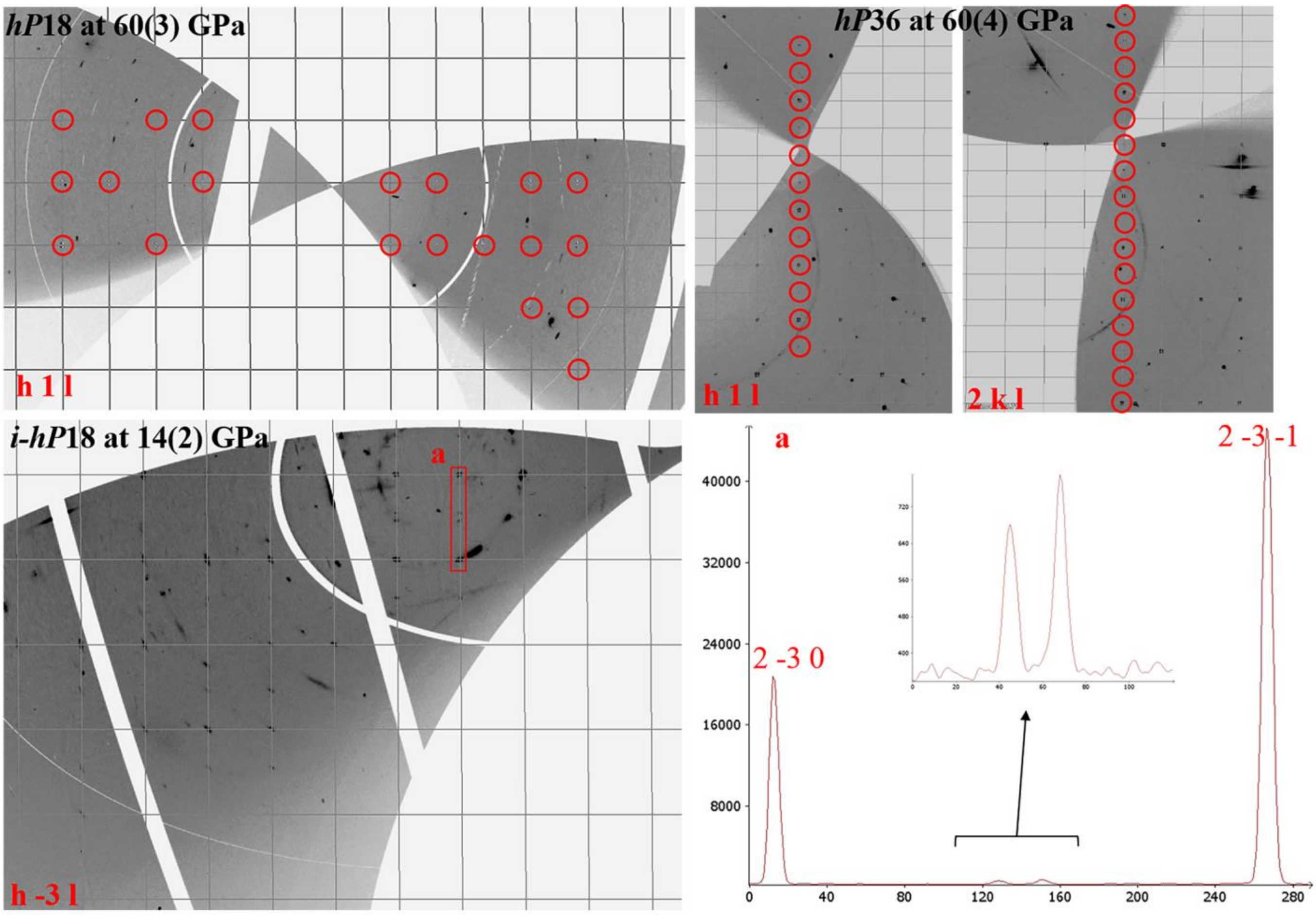

Fig. SD1. 3. Reconstructed reciprocal lattice planes of *hP*18 $Na_4Cl_5$ at 60(3) GPa, *hP*36 at 60(4) GPa, and *i-hP*18 at 14(2) GPa from the experimental SCXRD dataset using CrysAlis$^{Pro}$ software. One-dimensional line-profiles across the selected reflections a of *i-hP*18 are shown, highlighting the satellite peak splitting. The data was collected at ID15b beamline at ESRF with a beam size of 2×2 μm$^2$ for *hP*18 and *i-hP*18, and P02.2 beamline (beam size ~2.0 × 2.0 μm$^2$) of PETRA III for *hP*36 phase.

## Supplementary Discussion 2

Electronic properties of selected $Ga_4Ti_5$- and $Gd_5Si_3C$- structural type compounds at 1 bar

Nine compounds that crystallise in either the $Ga_4Ti_5$- or $Gd_5Si_3C$- structural type were taken from the Inorganic Crystal Structure Database (ICSD) for detailed electronic structure analysis. The materials were selected with the X-X contact within the $X_5$ chain had to be significantly shorter than the nearest-neighbour distance in the corresponding elemental metal.

1. $Ca_5Sb_3Cl$ (CSD 199808) has a Ca-Ca distance of 3.54 Å at 1 bar, which is ~0.41 Å shorter than that in a Ca metal (~3.95 Å). Full structural relaxation was carried out using the PBE functional, and after relaxation the Ca-Ca distance remains essentially unchanged at 3.54 Å. Band structure was calculated using the HSE06 functional, which suggest $Ca_5Sb_3Cl$ is a narrow-gap semiconductor with a bandgap of 0.27 eV. Spin-orbit-coupling was included in the band structure calculation.
2. $Hf_5Al_3C$ (CSD 199187) has a Hf-Hf distance of 2.84 Å at 1 bar, which is ~0.19 Å shorter than that in a Hf metal (~3.03 Å). Full structural relaxation was carried out using the PBE functional, and after relaxation the Hf-Hf distance remains essentially unchanged at 2.84 Å. Band structure was calculated using PBE, which suggest $Hf_5Al_3C$ is metallic at 1 bar with the contribution at the $E_F$ mainly from Hf.

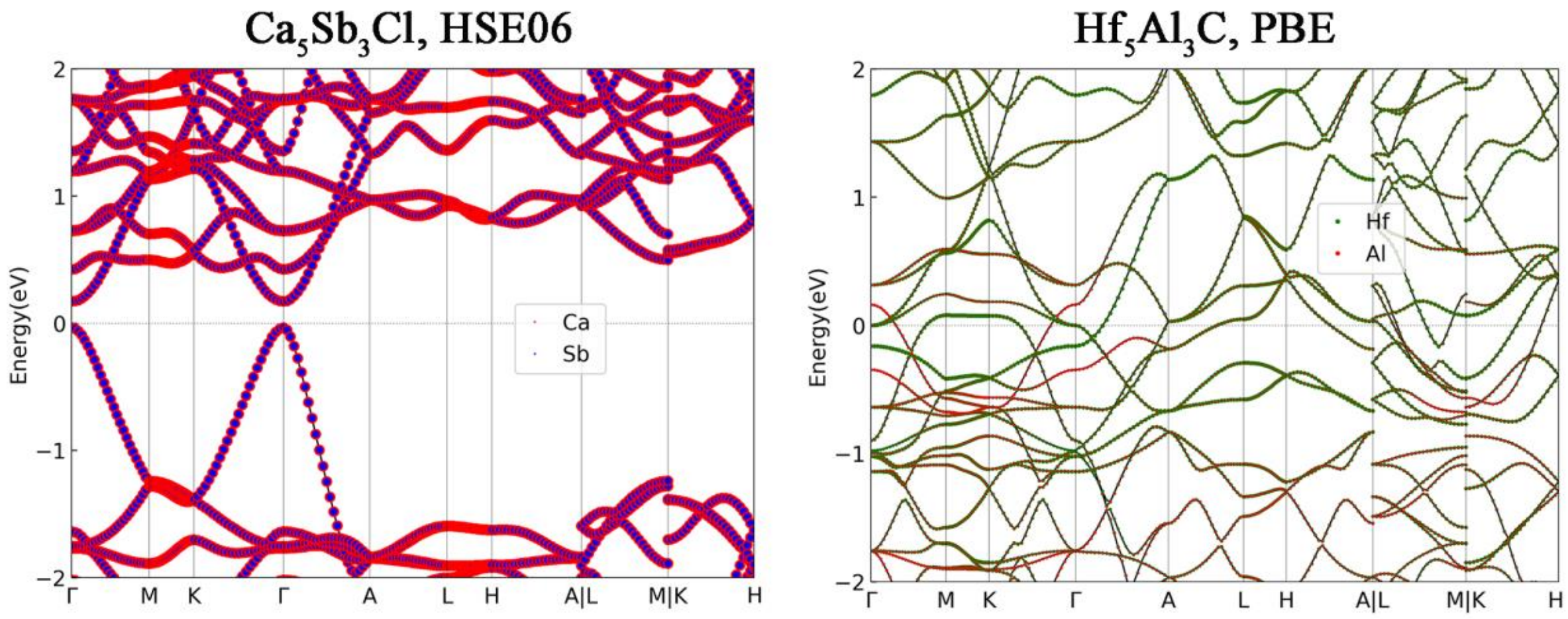

Fig. SD2. 1. Calculated band structure of $Ca_5Sb_3Cl$ and $Hf_5Al_3C$ at 1 bar. The Fermi energy was set at 0 eV.

3. $Mo_5Ge_3C$ (CSD 42922) has a Mo-Mo distance of 2.52 Å at 1 bar, which is ~0.20 Å shorter than that in a Hf metal (~2.73 Å). After full structural relaxation (using PBE) the Mo-Mo distance is 2.67 Å. Band structure was calculated using PBE, which suggest $Mo_5Ge_3C$ is metallic at 1 bar with the contribution at the $E_F$ mainly from Mo.
4. $Zr_5Sb_3O$ (CSD 100926) has a Zr-Zr distance of 2.85 Å at 1 bar, which is ~0.24 Å shorter than that in a Zr metal (~3.09 Å). After full structural relaxation (using PBE) the Zr-Zr distance is 2.92 Å. Band structure was calculated using PBE, which suggest $Zr_5Sb_3O$ is metallic at 1 bar with the contribution at the $E_F$ mainly from Zr.

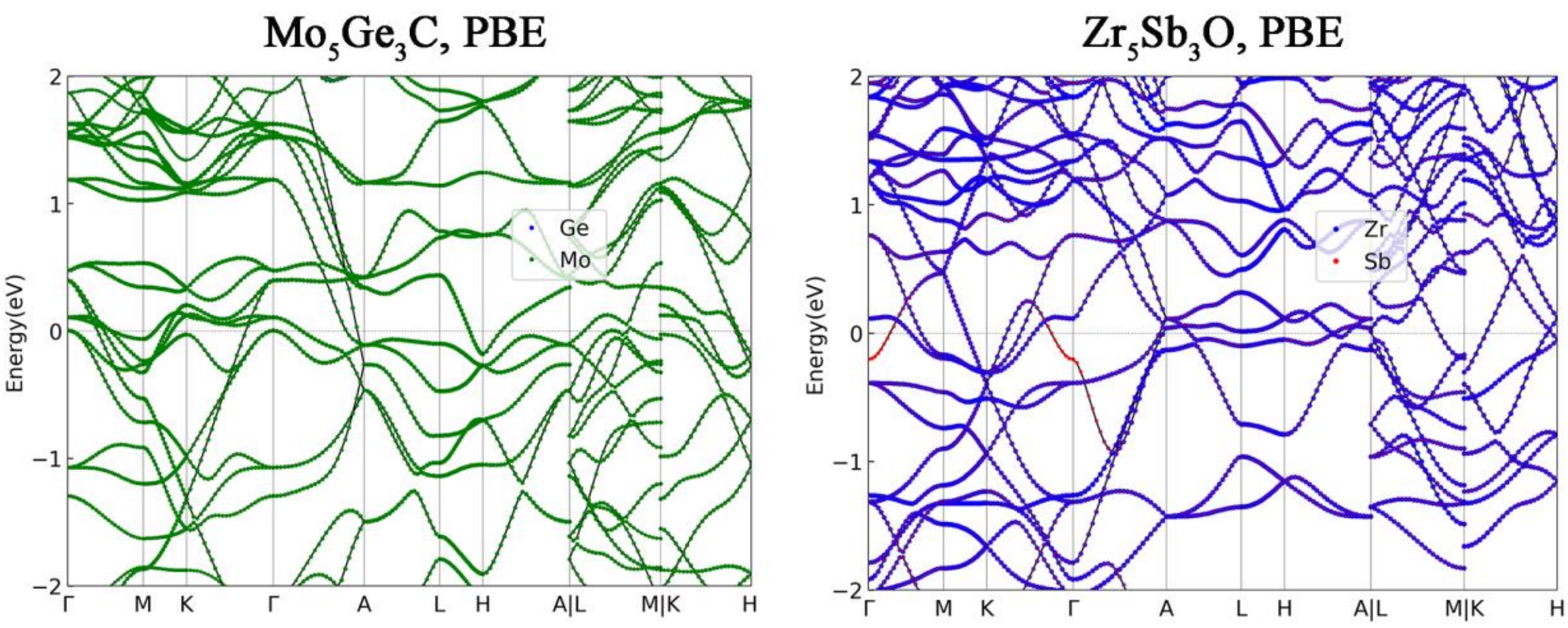

Fig. SD2. 2. Calculated band structure of $Mo_5Ge_3C$ and $Zr_5Sb_3O$ at 1 bar. The Fermi energy was set at 0 eV.

5. $Zr_5Al_4$ (CSD 197270) has a Zr-Zr distance of 2.91 Å at 1 bar, which is ~0.19 Å shorter than that in a Zr metal. After full structural relaxation (using PBE) the Zr-Zr distance is 2.90 Å. Band structure was calculated using PBE, which suggest $Zr_5Al_4$ is metallic at 1 bar with the contribution at the $E_F$ mainly from Zr.
6. $Y_5Si_3C$ (CSD 618792) has a Y-Y distance of 3.20 Å at 1 bar, which is ~0.36 Å shorter than that in a Y metal (~3.56 Å). After full structural relaxation (using PBE) the Y-Y distance is 3.21 Å. Band structure was calculated using PBE, which suggest $Y_5Si_3C$ is metallic at 1 bar with the contribution at the $E_F$ mainly from Y and Si.

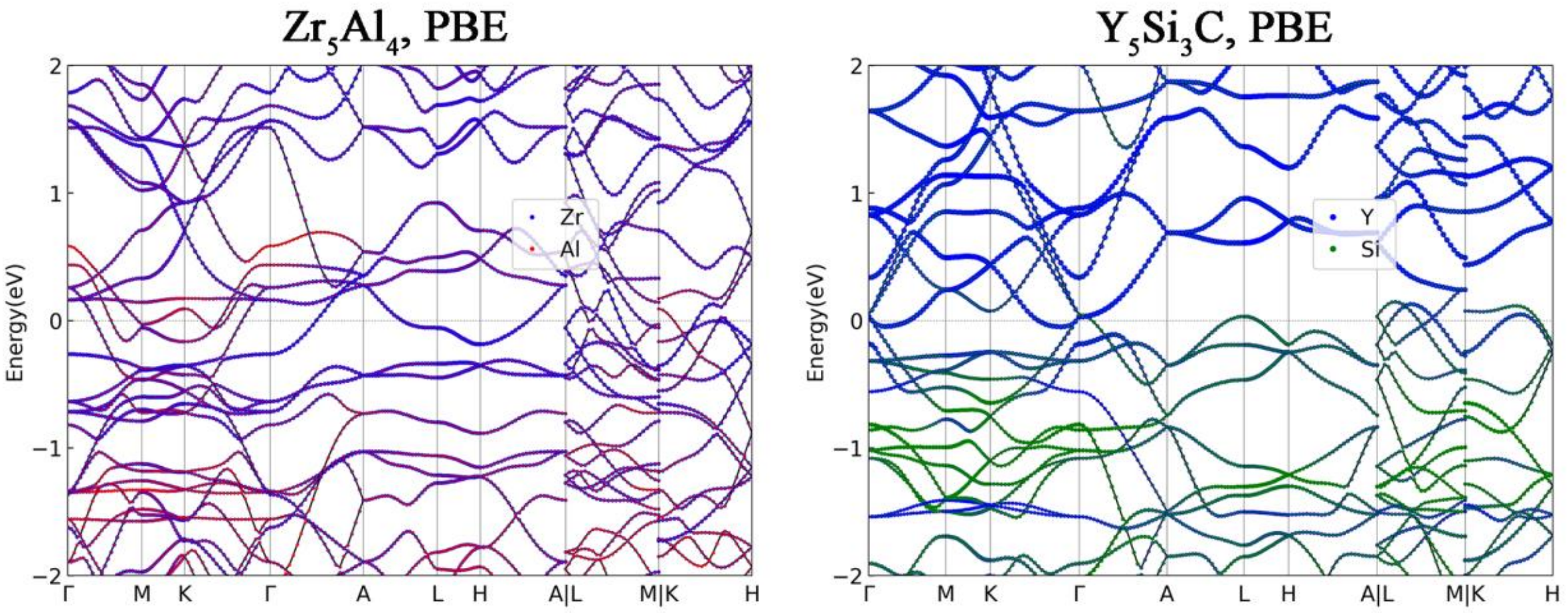

Fig. SD2 .3. Calculated band structure of $Zr_5Al_4$ and $Y_5Si_3C$ at 1 bar. The Fermi energy was set at 0 eV.

7. $Zr_5Si_3O$ (CSD 199821) has a Zr-Zr distance of 2.78 Å at 1 bar, which is ~0.31 Å shorter than that in a Zr metal. After full structural relaxation (using PBE) the Zr-Zr distance is 2.80 Å. Band structure was calculated using PBE, which suggest $Zr_5Si_3O$ is metallic at 1 bar with the contribution at the $E_F$ mainly from Zr.
8. $Zr_5Sn_3C$ (CSD 57056) has a Zr-Zr distance of 2.90 Å at 1 bar, which is ~0.19 Å shorter than that in a Zr metal. After full structural relaxation (using PBE) the Zr-Zr distance is 2.93 Å. Band structure was calculated using PBE, which suggest $Zr_5Sn_3C$ is metallic at 1 bar with the contribution at the $E_F$ mainly from Zr.

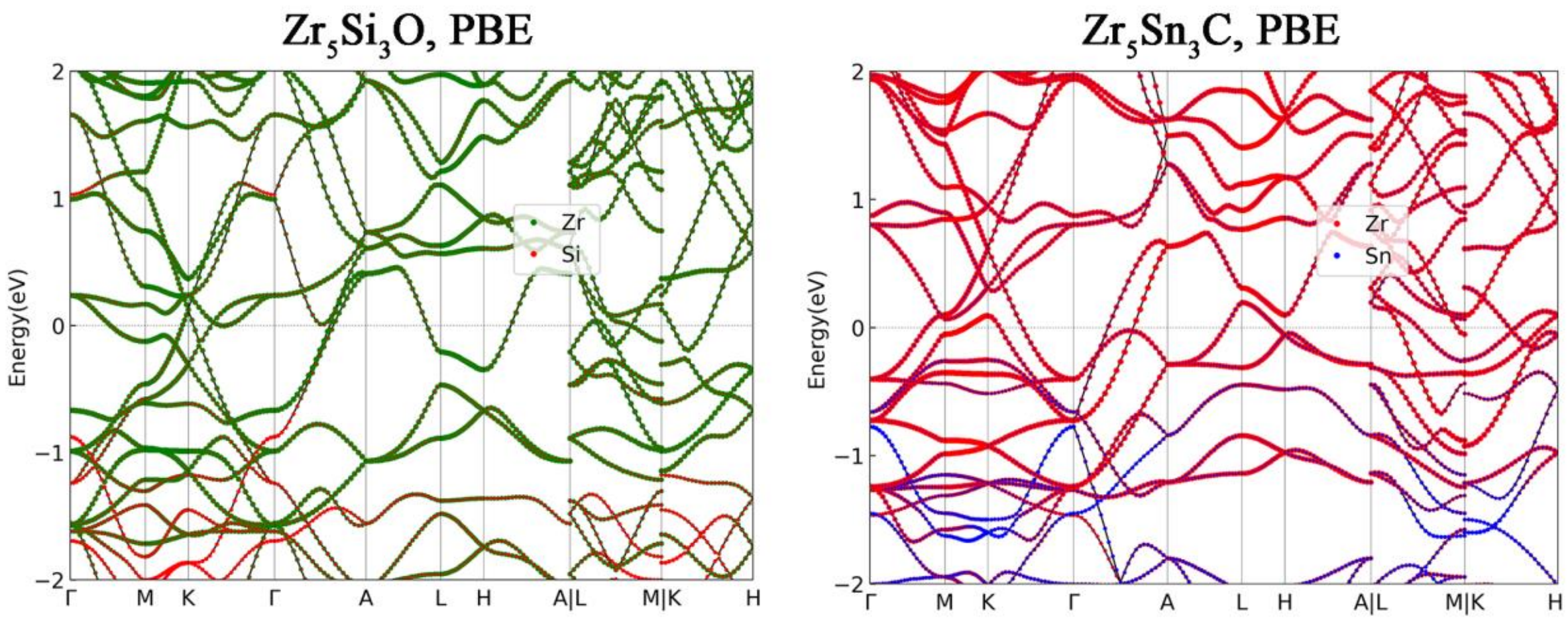

Fig. SD2. 4. Calculated band structure of $Zr_5Si_3O$ and $Zr_5Sn_3C$ at 1 bar. The Fermi energy was set at 0 eV.

9. A *hP*18-$Ba_3ScTe_5$ phase (CSD 120930) was also selected for a detailed electronic structure analysis (Fig. SD2. 5) with its band structure and eDOS do reveal q1D characteristics. However, according to Ref. (*26*), the *hP*18-$Ba_3ScTe_5$ phase is not stable under ambient conditions. Instead, $Ba_3ScTe_5$ crystalizes in an incommensurately modulated structure ($P\bar{6}(00\gamma)0$ and $\gamma = 0.3718(2)$) with the formation of long- and short-bonded pairs of Te atoms.

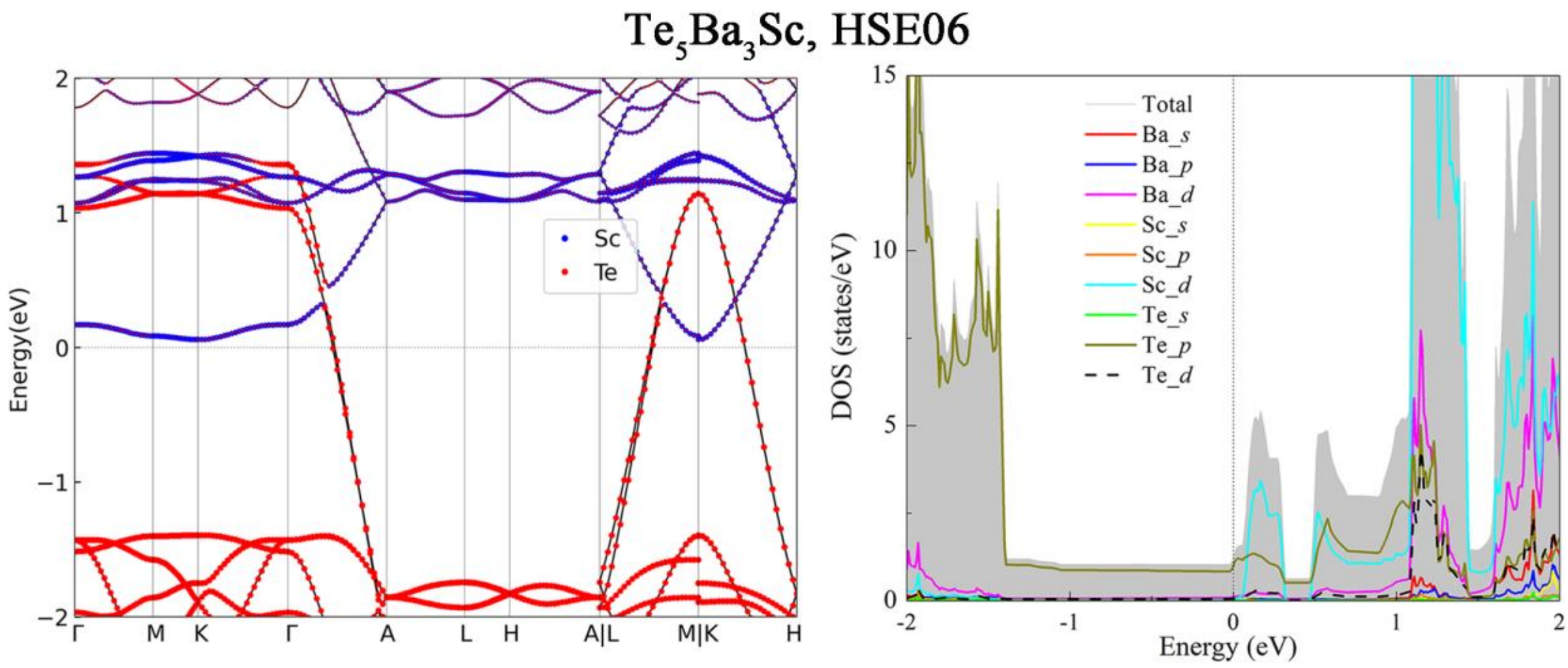

Fig. SD2. 5. Calculated band structure, and TDOS and PDOS curves of $Te_5Ba_3Sc$ at 1 bar. The Fermi energy was set at 0 eV.